\documentclass[journal,comsoc]{IEEEtran}
\usepackage[english]{babel}
\usepackage{booktabs}
\usepackage{comment}
\usepackage{mathrsfs}
\usepackage{amsmath}
\usepackage{graphicx}
\usepackage{amssymb}
\usepackage[colorlinks=true, allcolors=blue]{hyperref}
\usepackage{stfloats}     
\usepackage{color}    
\usepackage{enumitem}  
\usepackage{subcaption}
\usepackage{siunitx}                        
\usepackage{amsmath}                        
\usepackage{mathtools}                      
\usepackage{bm}                             
\usepackage{cite}
\usepackage{acronym}

\newacro{DoF}{degrees of freedom}
\newacro{4G}{fourth generation}
\newacro{6G}{sixth generation}
\newacro{BS}{base station}
\newacro{5G}{fifth generation}
\newacro{LIS}{large intelligent surface}
\newacro{SIS}{small intelligent surface}
\newacro{LoS}{line-of-sight}
\newacro{OFDM}{orthogonal frequency division multiplexing}
\newacro{WDM}{Wavenumber Division Multiplexing}
\newacro{CSI}{Channel State Information}
\newacro{SVD}{Singular Value Decomposition}
\newacro{MMSE}{Minimum Mean Square Error}
\newacro{PSWF}{Prolate Spheroidal Wave Function}
\newacro{MIMO}{Multiple-input-multiple-output}
\newacro{ULA}{uniform linear array}
\newacro{UPA}{uniform planar array}
\newacro{mmWave}{millimeter wave}
\newacro{THz}{terahertz}
\newacro{HoLoS}{holographic surface}
\newacro{SNR}{signal-to-noise ratio}
\newacro{UE}{user equipment}
\newacro{Wi-Fi}{wireless-fidelity}

\makeatletter
\newcommand{\mathleft}{\@fleqntrue\@mathmargin0pt}
\newcommand{\mathcenter}{\@fleqnfalse}
\makeatother

\title{Spatial Multiplexing in Near-Field\\ Line-of-Sight MIMO Communications: \\Paraxial and Non-Paraxial Deployments}

\author{Juan Carlos Ruiz-Sicilia,~\IEEEmembership{ Student Member,~IEEE,}
 Marco Di Renzo,~\IEEEmembership{  Fellow,~IEEE,}
 Placido Mursia,~\IEEEmembership{ Member,~IEEE,} 
 Aryan Kaushik,~\IEEEmembership{ Member,~IEEE,}
 Vincenzo Sciancalepore,~\IEEEmembership{ Senior Member,~IEEE  }
\thanks{Manuscript received March 15, 2024; revised May  15, 2024. J. C. Ruiz-Sicilia and M. Di Renzo are with Universit\'e Paris-Saclay, CNRS, CentraleSup\'elec, Laboratoire des Signaux et Syst\`emes, 3 Rue Joliot-Curie, 91192 Gif-sur-Yvette, France. (juan-carlos.ruiz-sicilia@centralesupelec.fr, marco.di-renzo@universiteparis-saclay.fr). P. Mursia and V. Sciancalepore are with NEC Laboratories Europe GmbH, 6G Networks, 69115 Heidelberg, Germany (name.surname@neclab.eu). A. Kaushik is with the School of Engineering \& Informatics, University of Sussex, UK (aryan.kaushik@sussex.ac.uk).}
\thanks{This work was supported in part by the European Commission through the H2020 MSCA 5GSmartFact project under grant agreement number 956670. The work of M. Di Renzo was supported in part by the European Commission through the Horizon Europe project titled COVER under grant agreement number 101086228, the Horizon Europe project titled UNITE under grant agreement number 101129618, and the Horizon Europe project titled INSTINCT under grant agreement number 101139161, as well as by the Agence Nationale de la Recherche (ANR) through the France 2030 project titled ANR-PEPR Networks of the Future under grant agreement NF-SYSTERA 22-PEFT-0006, and by the CHIST-ERA project titled PASSIONATE under grant agreement CHIST-ERA-22-WAI-04 through ANR-23-CHR4-0003-01. The work of P. Mursia was supported by the SNS JU 6G-DISAC Horizon Europe project under grant agreement 101139130. The work of V. Sciancalepore was supported by the Spanish Ministry of Economic Affairs and Digital Transformation, European Union NextGeneration-EU through the RISC-6G project under grant  agreement TSI-063000-2021-59.}
}

\begin{document}

\maketitle

\begin{abstract} 
Sixth generation (6G) wireless networks are envisioned to include aspects of energy footprint reduction (sustainability), besides those of network capacity and connectivity, at the design stage. This paradigm change requires radically new physical layer technologies. Notably, the integration of large-aperture arrays and the transmission over high frequency bands, such as the sub-terahertz spectrum, are two promising options. In many communication scenarios of practical interest, the use of large antenna arrays in the sub-terahertz frequency range often results in short-range transmission distances that are characterized by line-of-sight channels, in which pairs of transmitters and receivers are located in the (radiating) near field of one another. These features make the traditional designs, based on the far-field approximation, for multiple-input multiple-output (MIMO) systems sub-optimal in terms of spatial multiplexing gains. To overcome these limitations, new designs for MIMO systems are required, which account for the spherical wavefront that characterizes the electromagnetic waves in the near field, in order to ensure the highest spatial multiplexing gain without increasing the power expenditure. In this paper, we introduce an analytical framework for optimizing the deployment of antenna arrays in line-of-sight channels, which can be applied to paraxial and non-paraxial network deployments. In the paraxial setting, we devise a simpler analytical framework, which, compared to those available in the literature, provides explicit information about the impact of key design parameters. In the non-paraxial setting, we introduce a novel analytical framework that allows us to identify a set of sufficient conditions to be fulfilled for achieving the highest spatial multiplexing gain. The proposed designs are validated with numerical simulations.

\textbf{\textit{Index terms}} --- MIMO, near field, line-of-sight, spatial multiplexing, paraxial, non-paraxial settings.
\end{abstract}

\acresetall
\section{Introduction}\label{sec:intro}
\ac{MIMO} is a well-established technology, owing to its potential in boosting the rate of wireless networks by means of highly directional beamforming and spatial multiplexing capabilities, i.e., the possibility of serving many users in the same time-frequency resource with minimal interference~\cite{Zhang19}. Several current wireless communication systems, such as the \ac{4G}, the \ac{5G} and \ac{Wi-Fi}, as well as the upcoming \ac{6G}, rely on the \ac{MIMO} technology to fulfill their requirements in terms of throughput, reliability and multiple access~\cite{Akyildiz20}. In this context, the typical communication scenario considered in the literature consists of two multi-antenna transceivers deployed in a wireless channel characterized by the presence of strong multipath components, i.e., there exist several scattered paths between a multi-antenna transmitter and receiver, that effectively create a \ac{MIMO} channel with a high rank. As a result, by assuming coherent signal processing at the multi-antenna transmitter, i.e., by exploiting the available knowledge of the propagation channel, the system capacity scales linearly with the multiplexing gain, which, in turn, depends on the minimum number of antennas available at the transmitter and receiver \cite{Sanguinetti20}.

In recent years, in addition, the research community has shown considerable interest in exploiting high frequency bands, e.g., \ac{mmWave} and sub-\ac{THz} frequencies, for communication, due to the abundant availability of spectrum, which may potentially bring unprecedented gains in system performance~\cite{Wan21,Do2021}. At high frequency bands, wireless channels are characterized by a strong \ac{LoS} component, i.e., the direct link between a transmitter and a receiver is predominant, while the multipath becomes sparse as a result of material absorption and atmospheric attenuation~\cite{Wang20}. In mobile communications, \ac{LoS} links are often deemed to be low performing due to the low rank of the associated communication channel, which do not allow to effectively transmit multiple data streams concurrently, even if the transmitters and receivers are equipped with multi-antenna transceivers. This is because multi-antenna transceivers of moderate size that operate at low frequencies are typically designed to communicate at distances that exceed the Fraunhofer far-field distance, resulting in signals characterized by planar wavefronts~\cite{Liu23}.

At high frequencies, however, the signal power decays rapidly with the transmission distance, and hence the communication links are typically established over short distances and by relying on large antenna arrays. Therefore, multi-antenna transceivers of large size that operate at high frequencies do not necessarily operate at distances that exceed the Fraunhofer far-field distance, resulting in received signals that are characterized by spherical wavefronts. As a consequence, there is a renowned and increasing interest in near-field communications, i.e., network deployments in which the distance between the transmitters and receivers is shorter than the Fraunhofer far-field distance. Such short communication ranges offer opportunities for minimizing the energy expenditure of wireless systems, since the available power budget can be efficiently used while guaranteeing optimized system performance. Notably, the spherical wavefront that characterizes the electromagnetic waves can be leveraged for beam focusing, i.e., to concentrate the energy towards specified locations, in contrast to specified directions, as it is allowed through conventional beam steering designs obtained based on the far-field approximation~\cite{Liu23}. Due to the high energy concentration capability, beam focusing is viewed as an enabler to reduce the transmit power and the interference, paving the way towards energy sustainable communications~\cite{Huang19}. Under such conditions, conventional plane-wave approximations are not accurate anymore, as they lead to sub-optimal designs that do not provide the highest spatial multiplexing gain, thus deteriorating the spectral and energy efficiencies~\cite{Zhang23}.

For example, thanks to the large aperture of typical sub-\ac{THz} devices, which are characterized by a large number of antennas in a relatively small space, and the short transmission range of typical sub-\ac{THz} transmission links, recent works have shown that it is possible to effectively support multiple data streams, even in \ac{LoS} MIMO settings~\cite{Sanguinetti22}. However, this is only possible by appropriately modeling and exploiting the spherical wavefront of the transmitted signals in the near field~\cite{DiRenzo2023}. By relying on correct signal models, it is thus possible to design MIMO communication links that maximize the number of \ac{DoF} of a communication channel, i.e., the spatial multiplexing gain, while not increasing the amount of transmitted power and efficiently utilizing all the available radio frequency chains~\cite{Migliore06}, which inherently leads to energy sustainable physical layer designs for wireless communications.

\begin{table*}[!t]
\centering
\caption{Maximization of the DoF in LoS MIMO: Summary of related research works. The definitions of paraxial and non-paraxial deployments, and the models and approximations for the wavefront are given in Section \ref{subsec:PvsNP}.}
\label{tab:SoA}
\begin{tabular}{|l|l|l|l|l|l|}
\hline
\textbf{SNR regime} & \textbf{Array type} & \textbf{Array  orientation} & \textbf{Deployment} & \textbf{Wavefront approximation} & \textbf{References} \\ \hline
High           &  ULA       & Broadside           & Paraxial & Parabolic & \cite{Gesbert2002,Torkildson2011}           \\ \hline
High           &   UPA      & Broadside           & Paraxial & Parabolic & \cite{Larsson}                    \\ \hline
High           &   ULA      & Arbitrary           & Paraxial & Parabolic with projection & \cite{Bohagen2007ULA}             \\ \hline
High           &    UPA     & Arbitrary           & Paraxial & Parabolic with projection & \cite{Boehagen2007}               \\ \hline
Any            &   ULA      & Arbitrary           & Paraxial & Parabolic with projection & \cite{Do2020, Do2021a}            \\ \hline
Any            &   UPA      & Arbitrary           & Paraxial & Parabolic with projection & \cite{Do2021b}                    \\ \hline
High            &   UPA      & Arbitrary           & Paraxial and non-paraxial & Quartic & This paper                    \\ \hline
\end{tabular} 
\end{table*}

\subsection{State-of-the-art on Near-field LoS MIMO Communications} 
Recent works~\cite{Gesbert2002,Torkildson2011, Larsson, Bohagen2007ULA, Boehagen2007} have shown that \ac{MIMO} communication links in \ac{LoS} conditions (often referred to as \ac{LoS} \ac{MIMO}) can be appropriately optimized by considering near-field features. The proposed methods aim to decorrelate the received signals, by carefully designing the transmit and receive antenna arrays in terms of size and inter-element distances. In a nutshell, the overarching design criterion consists of optimizing the locations of the multiple antennas at the transmitters and receivers such that the resulting MIMO channel matrix has a full rank even in the absence of multipath propagation \cite{Do2021}. The relevance of optimizing the locations of the antenna elements in multi-antenna arrays is increasing even further lately, thanks to the development of new antenna technologies, including fluid antennas, movable antennas, and conformal metamorphic metasurfaces \cite{you2024generation}, \cite{mursia2024t3dris}. The methods and analysis presented next can be directly applied to these emerging antenna technologies, when the maximization of the DoF is the metric of interest.

A summary of state-of-the-art research works on the maximization of the DoF in LoS MIMO channels is given in Table \ref{tab:SoA}. In~\cite{Gesbert2002}, the authors identify the sensitivity of the rank of LoS \ac{MIMO} channels as a function of the antenna spacing. Moreover, it is shown that the orthogonality of the MIMO sub-channels under LoS conditions can only be ensured for short transmission ranges, i.e., in the near field. The design of near-field spatial multiplexing schemes for application to indoor \ac{mmWave} \ac{MIMO} channels is discussed in~\cite{Torkildson2011}. Specifically, the authors consider a \ac{ULA} with a constrained form factor, and they show that sparse array designs, i.e., when the inter-distances between the antenna-elements are larger than half of the wavelength, result in a spatially uncorrelated channel matrix, hence effectively providing the maximum number of spatial \ac{DoF}. The optimal inter-distance between the antenna arrays is identified, and it is shown to fulfill the Rayleigh criterion. In~\cite{Larsson}, it is shown that two communicating rectangular lattice antenna arrays, i.e., two \acp{UPA}, including rectangular arrays, square arrays, and \acp{ULA}, can achieve the maximum spatial multiplexing gain, provided that the inter-distance between the antenna elements is appropriately designed. The aforementioned papers consider a scenario in which the transmitting and receiving arrays are aligned in the broadside of each other. A more general analysis, which can be applied to geometrical models that account for any orientations of the transmitting and receiving arrays, is presented in \cite{Bohagen2007ULA} for \acp{ULA} and in \cite{Boehagen2007} for rectangular \acp{UPA}, respectively. Therein, the analysis is carried out by assuming the so-called \emph{parabolic} wavefront model (or approximation)~\cite{Do2023}, which can be applied in the near field provided that the transmission distance is much larger than the physical size of the transmitting and receiving arrays. In the literature, this is often referred to as the \emph{paraxial} setting. These research works assume that the LoS MIMO communication link operates at high \ac{SNR} values. In this operating regime, it is desirable to have a full-rank MIMO channel matrix with equal singular values, in order to maximize the spectral and energy efficiencies, assuming the same total consumed power. In the low \ac{SNR} regime, on the other hand, maximizing the received power is essential, and this scenario requires to beamform the transmitted signal over a channel whose maximum singular value is as large as possible. In \cite{Do2020}, the authors analyze the impact of the \ac{SNR} when optimizing LoS MIMO channels, and they show that, in order to strike a balance between spatial multiplexing and beamforming, the multi-antenna arrangements need to depend on the \ac{SNR}. Accordingly, the authors of \cite{Do2021a} propose to configure \acp{ULA} based on SNR-dependent rotations that maximize the rate. In \cite{Do2021b}, the same authors generalize the approach to \acp{UPA}, in order to reduce the array footprints.

\subsection{Paper Contributions} 
Against this background, we advance the state of art on modeling and optimizing LoS MIMO channels by introducing an analytical framework for optimizing the deployment of antenna arrays that can be applied to paraxial and non-paraxial deployments, i.e., when the transmission distance is not necessarily much larger than the physical size of the antenna arrays at the transmitter and receiver. Specifically, we provide two main contributions:
\begin{itemize} 
\item In the paraxial setting, we devise a simpler analytical framework than those available in the literature, e.g., in~\cite{Boehagen2007}, which provides explicit information about the impact of key design parameters, including the tilt and rotation of the antenna arrays. The proposed approach provides us with explicit analytical expressions for ensuring the highest spatial multiplexing gain with no restriction on the orientations and arrangements of the antenna arrays.
\item In the non-paraxial setting, we introduce a new analytical framework that allows us to identify the conditions that need to be fulfilled in order to achieve the highest spatial multiplexing gain in LoS MIMO channels. To gain design insights, we specialize the framework to MIMO deployments with linear arrays oriented in broadside. In this case, we propose an approximated analytical framework and introduce an optimized MIMO design that offers the largest spatial multiplexing gain, provided that the difference between the number of antennas at the two arrays is greater than a minimum value. Such a minimum number of excess antennas is estimated, in the considered case study, as well. To the best of our knowledge, there exist no contributions in the open technical literature that have tackled this design and optimization problem in non-paraxial settings.
\end{itemize} 

In terms of methodology, the proposed approach capitalizes on an approximation for spherical wavefronts that was recently introduced in \cite{ruizsicilia2023degrees}, which is referred to as the quartic wavefront approximation, and that can be applied to both the paraxial and non-paraxial settings. In \cite{ruizsicilia2023degrees}, however, the approach was applied to holographic MIMO channels and it has never been applied to LoS MIMO channels. Compared with the typical parabolic approximation for spherical wavefronts, the quartic approximation can be applied to a universal system of coordinates, hence resulting in simpler and more insightful analytical expressions in the paraxial setting, as well as enabling the analysis and optimization of LoS MIMO channels in non-paraxial settings.

\subsection{Paper Organization} 
The rest of the present paper is organized as follows. In Section~\ref{sec:SM}, the system model is introduced, the proposed analytical approach is presented, and the benefits with respect to currently available frameworks are discussed. In Section~\ref{sec:Parax}, the paraxial setting is considered and the proposed analytical approach is illustrated. In Section~\ref{sec:NonParax}, the approach is generalized for application to the non-paraxial setting, and the usefulness of the quartic approximation for spherical wavefronts is discussed. In Section~\ref{sec:NR}, extensive numerical results are illustrated to validate the proposed analytical frameworks, theoretical findings, and optimal design criteria for LoS MIMO channels. Finally, conclusions are drawn in Section~\ref{sec:Concl}.

{\textit{Notation}}: Bold lower and upper case letters represent vectors and matrices. $\mathbb{C}^{a \times b}$ denotes the space of complex matrices of dimensions $a \times b$. $(\cdot)^T$ denotes the transpose and $(\cdot)^*$ denotes the Hermitian transpose. $\mathbf{I}_N$ denotes the $N \times N$ identity matrix. $\mathbf{A}(i,k)$ denotes the $k$-th element of the $i$-th row of matrix $\mathbf{A}$. $j$ is the imaginary unit. $\mathcal{CN}(\mu,\sigma^2)$ denotes the complex Gaussian distribution with mean $\mu$ and variance $\sigma^2$. $\max \left\{ {a,b} \right\}$ returns the maximum between $a$ and $b$.

\section{System Model}\label{sec:SM}
We consider a \ac{MIMO} system, wherein the transmitter and receiver are \acp{UPA} with $L=L_1 L_2$ and  $M=M_1 M_2$ antenna elements, respectively, where $L_1$ ($M_1$) is the number of antenna elements in the first principal direction and $L_2$ ($M_2$) is the number of elements in the second principal direction. Without loss of generality, we assume that the larger array acts as the receiver, i.e., $M \ge L$. This corresponds to a typical uplink scenario. The transmitter lies on the $xz$-plane and is centered in $\mathbf{c}^{t}=(0,0,0)$, while the receiver is centered in $\mathbf{c}^{r}=(x_o,y_o,z_o)$. Hence, the distance between the center-points of the antenna arrays is
\begin{equation}
|\mathbf{c}^r - \mathbf{c}^t| = |\mathbf{c}_o|  = \sqrt{x_o^2 + y_o^2 + z_o^2} \,.
\end{equation}

The positions of the $l$-th transmit antenna element and the $m$-th receive antenna element are denoted by $\mathbf{r}_l^t = (x_l^t,y_l^t,z_l^t) + \mathbf{c}^t$ and $\mathbf{r}_m^r = (x_m^r,y_m^r,z_m^r) + \mathbf{c}^r$, where $a_l^t$ and $a_m^r$ are the local coordinates at the transmitter and receiver, respectively, for the $a$-axis with $a = \{x,y,z\}$.
%

The received signal $\mathbf{y} \in \mathbb{C}^{M \times 1}$ can be formulated as follows:
\begin{equation}
    \mathbf{y} = \mathbf{H} \mathbf{x} + \mathbf{n}
\end{equation}
where $\mathbf{n}$ is the additive white Gaussian noise at the receiver, with $\mathbf{n} \sim \mathcal{CN}(0, \sigma^2\mathbf{I}_M)$, $\mathbf{x} \in \mathbb{C}^{L \times 1}$ is the transmitted signal, and $\mathbf{H} \in \mathbb{C}^{M \times L}$ is the channel matrix from the multi-antenna transmitter to the multi-antenna receiver.

According to the considered system model, the channel capacity is given as follows \cite{Tse}:
\begin{equation}
\label{eq:capacity}
    C = \sum_{i=1}^R \log_2 \left(1 + \frac{P_i}{\sigma^2} \lambda_i\right)
\end{equation}
where $\lambda_i$ is the $i$-th largest eigenvalue of $\mathbf{G} = \mathbf{H}^{*} \mathbf{H}$, $R \le L$ is the rank of $\mathbf{G}$,  and $P_i$ is the power allocated to the $i$-th communication mode. The power allocation is subject to the constraint $\sum_{i=1}^{R} P_i = P_T$, where $P_T$ is the maximum power budget at the transmitter. 

In general, only some of the available communication modes have a significant coupling intensity $\lambda_i$, and hence are valuable for communication. Consequently, the optimal power allocation policy, i.e., the waterfilling algorithm \cite{Tse}, allocates little or no power to the weakly-coupled modes. A metric to measure the number of effective communication modes is the effective rank \cite{effRank}. This metric, which is denoted by $N_{\mathrm{eff}}$, is bounded by $N_{\mathrm{eff}} \in [1, R]$. The effective rank attains the upper bound $R$ when all the eigenvalues $\lambda_i$ are equal, and it attains the lower bound when only one eigenvalue is non-zero. 

In the high \ac{SNR} regime, which is typical in \ac{LoS} conditions, as considered in this paper, the capacity is maximized when the rank of $\mathbf{G}$ is $R=L$ and the $R$ eigenvalues have the same magnitude, which implies that $N_{\mathrm{eff}} = L$ and the equipower allocation policy is optimal, i.e., $P_i = P_T/L$. This condition is ensured when $\mathbf{H}$ is an orthogonal matrix, and hence $\mathbf{G} = \lambda_0 \mathbf{I}_L$ with $\lambda_0$ being the magnitude of all the eigenvalues. Accordingly, we need to ensure that the antenna elements of the arrays are placed at locations that fulfill the following condition:
\begin{equation}
\label{eq:orthogoCondGen}
    \mathbf{G}(u,v) = \sum_{m=1}^M \left[ \mathbf{H}(m,u)\right]^* \mathbf{H}(m,v) = 0 \quad \forall u \not= v = 1, 2 \ldots, L
\end{equation}
which ensures that the matrix $\mathbf{H}$ is, by definition, orthogonal.

The effective rank is an appropriate figure of merit to estimate the rank of $\mathbf{G}$ and to evaluate the similarity among the eigenvalues of $\mathbf{G}$, since it attains its upper bound when all the eigenvalues are equal. Hence, it can be utilized as the metric to verify the accomplishment of \eqref{eq:orthogoCondGen}, by means of numerical simulations. Next, we devise analytical expressions for ensuring that \eqref{eq:orthogoCondGen} is fulfilled and, in Section~\ref{sec:NR}, we validate them against the effective rank.

As far as the channel model is concerned, we assume a free-space \ac{LoS} propagation channel. Accordingly, the link from the $l$-th antenna of the MIMO transmitter to the $m$-th antenna of the MIMO receiver can be modeled as \cite{Miller2000}
\begin{equation}
\label{eq:channelMatrix}
    \mathbf{H}(m,l) = \frac{e^{j k_0 |\mathbf{r}_m^r - \mathbf{r}_l^t|}}{4\pi |\mathbf{r}_m^r - \mathbf{r}_l^t|}
\end{equation}
where $k_0 = 2 \pi/\lambda$ and $|\mathbf{r}_m^r - \mathbf{r}_l^t|$ is the distance between the antenna elements.

\subsection{Paraxial and Non-Paraxial Settings}\label{subsec:PvsNP}
\begin{figure}[!t]
     \centering
    \includegraphics[width=0.8\linewidth]{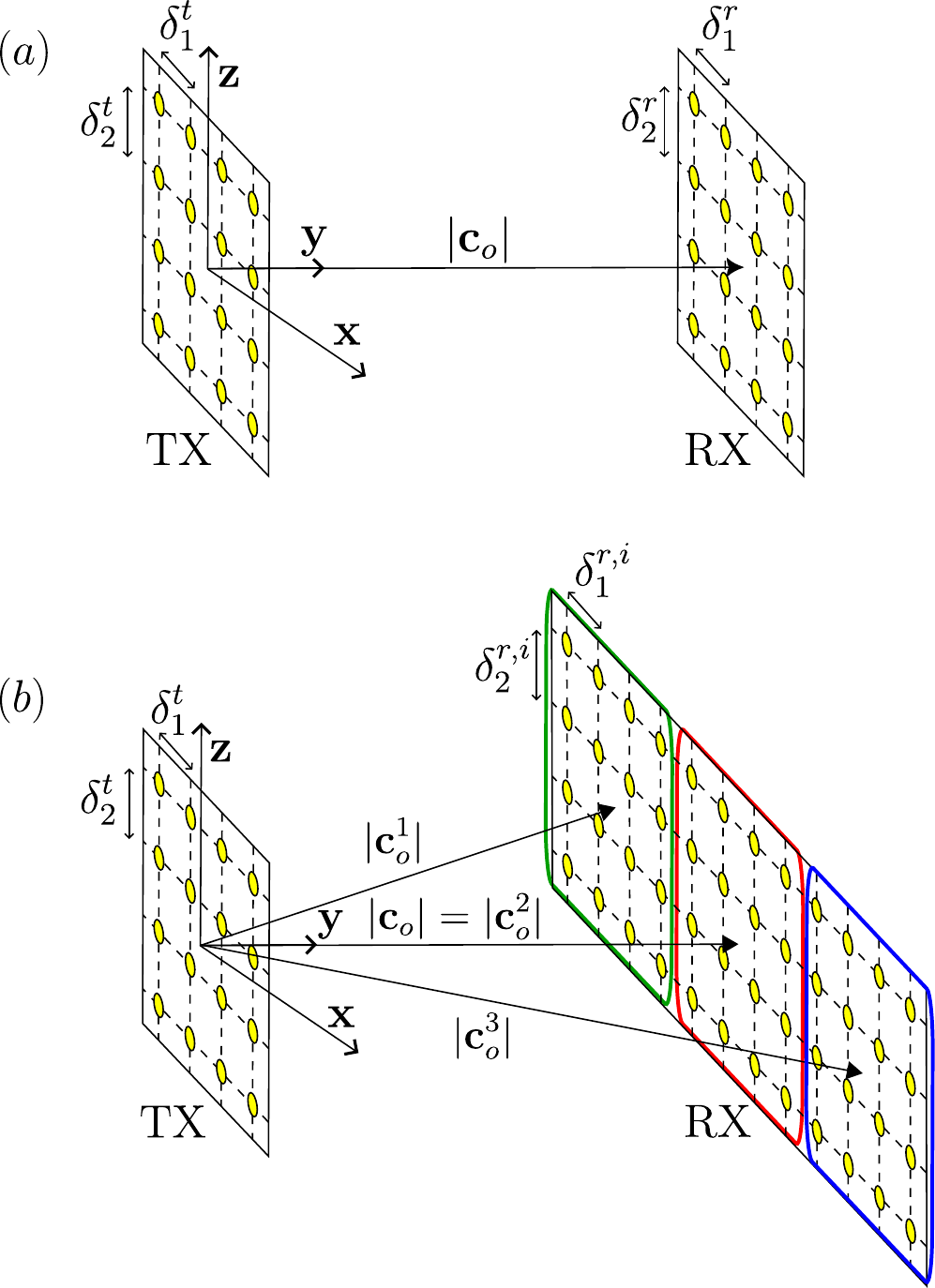}
    \caption{Paraxial (a) and non-paraxial (b) settings.}
    \label{fig:paraxvsnonparax} 
\end{figure}

According to \eqref{eq:channelMatrix}, the signal emitted by the antenna array at the transmitter has a spherical wavefront. When the physical apertures of the transmitter and receiver, the signal wavelength, and the transmission  distance between their center-points fulfill the Fraunhofer far-field condition, the MIMO system operates in the far field, and the wavefront of the received signal can be well approximated as planar. This implies that the columns of the channel matrix $\mathbf{H}$ are correlated, and that, regardless of the arrangements of the antenna elements, $\mathbf{G}$ has a unit rank, i.e., $N_{\mathrm{eff}} = 1$. 

If the antenna arrays are located closer than the Fraunhofer far-field distance, the planar wavefront approximation is not accurate anymore. However, there exists a region where the apertures of the antenna arrays are still small compared with the distance between their center-points but the wavefront of the received signal is spherical. This occurs when the size of the antenna arrays is sufficiently small as compared with the distance between their center points. In mathematical terms, this condition can be formulated as follows:
\begin{equation}
\label{eq:paraxAssum}
    x^t_l, y^t_l, z^t_l, x^r_m, y^r_m, z^r_m \ll |\mathbf{c}_o| \, .
\end{equation}

This deployment scenario is illustrated in Fig. \ref{fig:paraxvsnonparax}(a), and it is referred to as the paraxial setting. Figure \ref{fig:paraxvsnonparax}(a) shows the antenna arrays in the broadside configuration, but the condition in \eqref{eq:paraxAssum} is independent of the tilt and rotation of the antenna arrays with respect to one another. In the paraxial setting, the channel model in \eqref{eq:channelMatrix} can be simplified to make it more tractable, as detailed in \cite{Boehagen2007,Do2023}. Therein, the authors utilize an approach that consists of (i) changing the system of coordinates so that the centers of the two antenna arrays are aligned along the axis connecting their center-points and (ii) projecting the arrays onto the plane that is perpendicular to the axis that connects their center-points. By utilizing this change of reference system, the conventional parabolic approximation for spherical wavefronts can be applied and is sufficiently accurate \cite{Do2023}.

In \cite{ruizsicilia2023degrees}, the authors propose an alternative approach based on a quartic approximation for the spherical wavefront. This approach is independent of the system of coordinates being considered. It utilizes a simple parametrization to represent the antenna arrays, which is not based on projections of the antenna arrays, and that leads to a more insightful problem formulation and understanding of the obtained analytical framework. The methods of analysis proposed in \cite{Boehagen2007,Do2023} and in \cite{ruizsicilia2023degrees} are equivalent in the paraxial setting. This is further elaborated in Section \ref{sec:Parax}.

By contrast, when the antenna arrays are not in the paraxial setting, i.e., \eqref{eq:paraxAssum} is not fulfilled, none of the two approaches can be applied as originally reported in~\cite{Boehagen2007,Do2023} and~\cite{ruizsicilia2023degrees}. An example of non-paraxial setting is shown in Fig. \ref{fig:paraxvsnonparax}(b). The large antenna array may represent a distributed multi-antenna base station that communicates with a multi-antenna user equipment, when the size of the base station and user equipment are much larger and much smaller than the distance between their center-points $|\mathbf{c}_o|$, respectively (uplink). We see next that the optimal placements for the antennas of the base station have inter-distances larger than half of the wavelength.

While none of the approaches reported in \cite{Boehagen2007,Do2023} and \cite{ruizsicilia2023degrees} can be applied directly, the method based on the quartic approximation proposed in \cite{ruizsicilia2023degrees} can be generalized for analyzing the non-paraxial setting, since it is based on an independent system of coordinates without the need for applying projections. Specifically, we tackle the problem at hand by partitioning the large antenna array in Fig. \ref{fig:paraxvsnonparax}(b) into smaller sub-arrays. Each sub-array is chosen to be small enough for ensuring that the paraxial approximation holds true in a system of coordinates that is common to all the sub-arrays. Under these assumptions, the quartic approximation method can be applied to the channel between the multi-antenna transmitter and each sub-array of the multi-antenna receiver. On the other hand, the approach based on projections cannot be directly applied to the proposed method of analysis based on sub-arrays, as it is not possible to align a single system of coordinates with the many axis that connect the center of each receiving sub-array with the center of the multi-antenna transmitter. It is worth mentioning that the notion of array of sub-arrays can be found in the literature \cite{Torkildson2011, Lin16, Do2020}, but the proposed approach is different, since (i) we utilize the decomposition in sub-arrays for modeling the non-paraxial setting and (ii) we optimize the locations of the antenna elements in each sub-array. In \cite{Torkildson2011, Lin16, Do2020}, the paraxial setting is analyzed and the antenna elements in each sub-array are spaced by half of the wavelength.

In Section \ref{sec:Parax}, we analyze the conditions for achieving channel orthogonality when both the transmitter and the receiver are deployed in the paraxial setting, by using the quartic approximation. In Section \ref{sec:NonParax}, we derive the conditions for ensuring channel orthogonality when the receiver is large enough that the paraxial approximation is not fulfilled anymore, by combining the quartic approximation with the sub-array partitioning approach.

\section{Paraxial Setting}\label{sec:Parax} 
In this section, we identify the conditions to make the channel matrix $\mathbf{H}$ orthogonal, i.e., to fulfill \eqref{eq:orthogoCondGen}, in the case of paraxial setting. First, we introduce the channel model based on the paraxial approximation and then we analyze the orthogonality condition in \eqref{eq:orthogoCondGen}.

\subsection{Paraxial Channel Model}\label{subsec:ParaxChannel}
\begin{figure}[t]
     \centering
\includegraphics[width=1\linewidth] {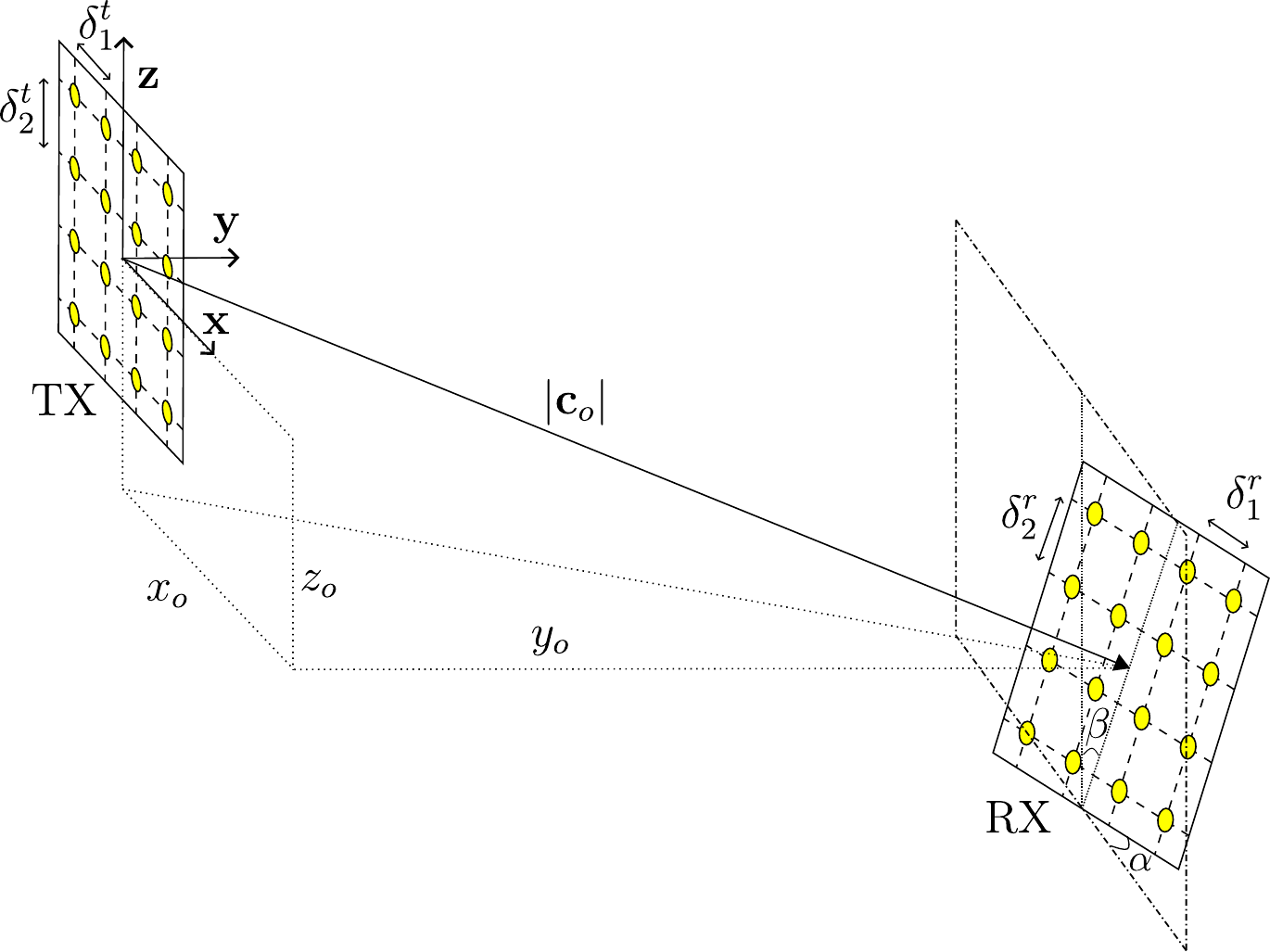}
    \caption{Coordinate system for the paraxial setting.}
    \label{fig:paraxSetup} 
\end{figure}
Under the assumption in \eqref{eq:paraxAssum}, the amplitude of \eqref{eq:channelMatrix} changes slowly and it can be approximated as a constant over the MIMO receiver, i.e., $|\mathbf{r}_m^r - \mathbf{r}_l^t| \approx |\mathbf{c}_o|$. By contrast, the phase in \eqref{eq:channelMatrix} is very sensitive to the variations of $|\mathbf{r}_m^r - \mathbf{r}_l^t|$. By denoting $\sum {f\left( a \right)}  = f\left( x \right) + f\left( y \right) + f\left( z \right)$, the distance $|\mathbf{r}_m^r - \mathbf{r}_l^t|$ in the phase term can be approximated as
\begin{align}
    |\mathbf{r}_m^r - \mathbf{r}_l^t| &= \sqrt{\sum \nolimits (a^r_m + a_o - a^t_l)^2}  \\
    &= |\mathbf{c}_o| \sqrt{1 +  \frac{\rho\left(\mathbf{r}_l^t, \mathbf{r}_m^r\right)}{|\mathbf{c}_o|^2}} \\
    & \approx |\mathbf{c}_o| \left[1 + \frac{\rho\left(\mathbf{r}_l^t, \mathbf{r}_m^r\right)}{2|\mathbf{c}_o|^2} - \frac{ \rho^2 \left(\mathbf{r}_l^t, \mathbf{r}_m^r\right)}{8|\mathbf{c}_o|^4} \right] \label{eq:2nTaylorApprox}
\end{align}
where 
\begin{align}
    \rho\left(\mathbf{r}_l^t, \mathbf{r}_m^r\right) &= 2\sum \nolimits a_o (a^r_m - a^t_l) + \sum \nolimits (a^r_m - a^t_l)^2 \label{eq:rhoExact}\\
    & \approx 2\sum \nolimits a_o (a^r_m - a^t_l) . \label{eq:rhoApprox}
\end{align}

The approximation in \eqref{eq:2nTaylorApprox} stems from Taylor's approximation $\sqrt{1+t} \approx 1 + t/2 - t^2/8$. Also, the approximation in \eqref{eq:rhoApprox} can be applied when $2\sum \nolimits a_o (a^r_m - a^t_l) \gg \sum \nolimits (a^r_m - a^t_l)^2$, i.e., when the misalignment between the centers of the arrays is larger than the size of the arrays. Hence, \eqref{eq:rhoApprox} is suitable for characterizing non-broadside settings, but it can not be applied in broadside settings, given that $|a_o| \ll 1$, for at least two Cartesian coordinates, in this latter case.

Compared with the typical parabolic approximation utilized in the paraxial setting~\cite{Boehagen2007,Do2023}, the proposed approximation in \eqref{eq:2nTaylorApprox} utilizes an additional term of the Taylor expansion. This is because, as mentioned, the system of coordinates does not coincide with the axis connecting the center-points of the antenna arrays and no projections onto this axis are applied. More specifically, the first-order term in \eqref{eq:2nTaylorApprox} is dominant in broadside deployments while the second-order term needs to be added in non-broadside deployments. For this reason, the exact formulation of $\rho\left(\mathbf{r}_l^t, \mathbf{r}_m^r\right)$ in \eqref{eq:rhoExact} needs to be utilized to compute the first-order term in \eqref{eq:2nTaylorApprox}, while the approximated formulation in \eqref{eq:rhoApprox} is sufficient to compute the second-order term  in \eqref{eq:2nTaylorApprox}. In mathematical terms, \eqref{eq:2nTaylorApprox} can be approximated as follows:
\begin{align}
    |\mathbf{r}_m^r - \mathbf{r}_l^t| & \approx |\mathbf{c}_o| \left[1 + \frac{\rho\left(\mathbf{r}_l^t, \mathbf{r}_m^r\right)}{2|\mathbf{c}_o|^2} - \frac{ \rho^2 \left(\mathbf{r}_l^t, \mathbf{r}_m^r\right)}{8|\mathbf{c}_o|^4} \right] \nonumber  \\
& \approx |\mathbf{c}_o| \left[1 + \frac{2\sum \nolimits a_o (a^r_m - a^t_l) + \sum \nolimits (a^r_m - a^t_l)^2}{2|\mathbf{c}_o|^2} 
\vphantom{\frac{\left(\sum \nolimits a_o (a^r_m - a^t_l)\right)^2}{2|\mathbf{c}_o|^4}}\right. \nonumber\\ 
& \left.\qquad \quad - \frac{\left(\sum \nolimits a_o (a^r_m - a^t_l)\right)^2}{2|\mathbf{c}_o|^4} \right] \, .
\label{eq:QuarticExplicit}
\end{align}

The modeling approach in \eqref{eq:QuarticExplicit} is referred to as the quartic approximation for the wavefront in \cite{ruizsicilia2023degrees}. Based on the quartic approximation, the channel matrix in \eqref{eq:channelMatrix} can be expressed as follows:
\begin{equation}
\label{eq:quarticApprox}
    \mathbf{H} \approx \frac{1}{4\pi|\mathbf{c}_o|} \mathbf{F}_{\mathrm{RX}} \mathbf{P} \mathbf{F}_{\mathrm{TX}}^{*}
\end{equation}
where $\mathbf{F}_{\mathrm{TX}} \in \mathbb{C}^{L \times L}$ and $\mathbf{F}_{\mathrm{RX}} \in \mathbb{C}^{M \times M}$ are diagonal matrices and $\mathbf{P} \in \mathbb{C}^{M \times L}$ is a non-diagonal matrix, which are defined as follows:
\mathleft
\begin{align}
\label{eq:focFunctionTX}
    \mathbf{F}_{\mathrm{TX}}(l,l) = \exp \Bigg\{ & j \frac{k_0}{2|\mathbf{c}_o|}\left[ (x^t_l)^2 + (y^t_l)^2 + (z^t_l)^2  - 2 x_o x^t_l
    \vphantom{\frac{(z^t_l)^2}{|\mathbf{c}_o|^2}}\right. \nonumber\\ 
    & \left. \hspace{-0.5cm}  - 2 y_o y^t_l - 2 z_o z^t_l - \frac{(x_o x^t_l + z_o z^t_l)^2}{|\mathbf{c}_o|^2}\right]\Bigg\} 
\end{align}
\begin{align}
\label{eq:focFunctionRX}
    \mathbf{F}_{\mathrm{RX}}(m,m) = \exp \Bigg\{ & j \frac{k_0}{2|\mathbf{c}_o|}\left[ (x^r_m)^2 + (y^r_m)^2 + (z^r_m)^2  + 2 x_o x^r_m
    \vphantom{\frac{(z^t_l)^2}{|\mathbf{c}_o|^2}}\right. \nonumber\\ 
    & \left. \hspace{-0.5cm} + 2 y_o y^r_m + 2 z_o z^r_m - \frac{(x_o x^r_m + z_o z^r_m)^2}{|\mathbf{c}_o|^2}\right]\Bigg\}
\end{align}
\begin{align}
\label{eq:P_Parax}
    \mathbf{P}(m,l) = \exp \Bigg\{& -j \frac{k_0}{|\mathbf{c}_o|} \left[ x^r_m x^t_l + y^r_m y^t_l + z^r_m z^t_l 
     \vphantom{\frac{(z^t_l)^2}{|\mathbf{c}_o|^2}} \right.  \nonumber\\
    & \left. \hspace{-1.5cm} - \frac{(x_o x^r_m + y_o y^r_m + z_o z^r_m ) (x_o x^t_l + y_o y^t_l + z_o z^t_l) }{|\mathbf{c}_o|^2}\right]\Bigg\} \, .
\end{align}
\mathcenter
and the off-diagonal elements of $\mathbf{F}_{\mathrm{TX}}$ and $\mathbf{F}_{\mathrm{RX}}$ are equal to zero, as they are diagonal matrices.

\subsection{Channel Orthogonality}\label{subsec:OrthoParax}
According to \eqref{eq:quarticApprox}, the matrix $\mathbf{G}$ can be written as follows:
\begin{equation}
\label{eq:GParax}
    \mathbf{G} = \frac{1}{(4\pi|\mathbf{c}_o|)^2} \left[ \mathbf{F}_{\mathrm{RX}} \mathbf{P} \mathbf{F}_{\mathrm{TX}}^{*}\right]^* \mathbf{F}_{\mathrm{RX}} \mathbf{P} \mathbf{F}_{\mathrm{TX}}^{*} \, .
\end{equation}

Given that $\mathbf{F}_{\mathrm{RX}}^*\mathbf{F}_{\mathrm{RX}} = \mathbf{I}_M$, \eqref{eq:GParax} can be simplified as
\begin{equation}
\label{eq:GParax2}
    \mathbf{G} = \frac{1}{(4\pi|\mathbf{c}_o|)^2} \mathbf{F}_{\mathrm{TX}} \mathbf{P}^* \mathbf{P} \mathbf{F}_{\mathrm{TX}}^{*} \, .
\end{equation}

Since $\mathbf{F}_{\mathrm{TX}}$ is a diagonal matrix, it does not have any impact on the diagonalization of $\mathbf{G}$. Therefore, the orthogonality condition in \eqref{eq:orthogoCondGen} can be expressed as follows:
\begin{equation}
\label{eq:orthogoCondParax}
    \sum_{m=1}^M \left[ \mathbf{P}(m,u)\right]^* \mathbf{P}(m,v) = 0 \quad \forall u \ne v
\end{equation}
where $u,v=1, 2, \ldots, L$ denote two generic antenna elements of the MIMO transmitter.

To proceed further, we consider the following parametrization for the $u$-th and $v$-th antenna elements of the array at the transmitter, and for the $m$-th antenna element of the array at the receiver, respectively (see Fig. \ref{fig:paraxSetup}):
%
%
\begin{align}
\label{eq:paramTx}
    & \mathbf{r}_u^{t} = \left(u_1^c \delta_1^{t} , 0 ,u_2^c \delta_2^{t} \right), \qquad \mathbf{r}_v^{t} = \left(v_1^c \delta_1^{t} , 0 ,v_2^c \delta_2^{t} \right)
\end{align}
\begin{align}
\label{eq:paramRXParax}
    \mathbf{r}_m^{r} = &\left( \delta_1^{r} m_1^c \cos \alpha - \delta_2^{r} m_2^c \sin \beta \sin \alpha, \right. \nonumber\\ 
    & \hspace{0.25cm}\left. \delta_1^{r} m_1^c \sin \alpha  + \delta_2^{r} m_2^c \sin \beta \cos \alpha, \delta_2^{r} m_2^c \cos\beta \right) + \mathbf{c}_o
\end{align}
with
\begin{align}
\label{eq:centeringCoor}
 & u_a^c = u_a -\frac{L_a-1}{2}, \qquad 
 v_a^c = v_a -\frac{L_a-1}{2}, \\ 
 & m_a^c = m_a -\frac{M_a-1}{2}
\end{align}
where $u = (u_1-1)L_1+u_2$, $v = (v_1-1)L_1+v_2$, and $m = (m_1-1)M_1+m_2$, with $u_a = 1, 2, \ldots, L_a$ and $v_a = 1, 2, \ldots, L_a$ for $a=\{1,2\}$ denote the indices of the $u$-th and $v$-th transmit antenna element along the first ($a=1$) and second ($a=2$) principal directions, respectively. Moreover, $m_1 = 1, 2, \ldots, M_1$ and $m_2 = 1, 2, \ldots, M_2$ denote the indices of the $m$-th  receive antenna element along the first and second principal directions, respectively. Also, $\delta_1^{t}$ ($\delta_1^{r}$) and $\delta_2^{t}$ ($\delta_2^{r}$) denote the inter-distances between the antenna elements in the first and second principal directions, respectively. Lastly, the angles $\alpha$ and $\beta$ denote a rotation and a tilt with respect to the $x$-axis and the $z$-axis, respectively. Based on the considered parametrization, the positions of the antenna elements within the arrays are determined based on their inter-distances. Since the inter-distance is kept fixed for all the elements, the resulting design corresponds to a uniform array.

Applying the parametrization in \eqref{eq:GParax} and \eqref{eq:GParax2} to the matrix $\mathbf{P}$, it can be rewritten as
\begin{align}
\label{eq:P_Param}
    \mathbf{P}(m,u) = \exp \bigg\{- j \frac{k_0}{|\mathbf{c}_o|} \left[ (\tau_{11} \delta_1^t u_1^c + \tau_{12} \delta_2^t u_2^c) \delta_1^r m_1^c \right. \nonumber \\ \left.
    + (\tau_{21} \delta_1^t u_1^c + \tau_{22} \delta_2^t u_2^c) \delta_2^r m_2^c \right] \bigg\}
\end{align}
where
\begin{align}
&\tau_{11} = \cos \alpha - x_o \tau_1, \quad &\tau_{12} &= - z_o\tau_1 \\
&\tau_{21} = - \sin \beta \sin \alpha - x_o\tau_2, \quad &\tau_{22} &= \cos \beta - z_o\tau_2 \label{eq:tau_2} 
\end{align}
with $\tau_1 = (x_o \cos \alpha + y_o \sin \alpha)|\mathbf{c}_o|^{-2}$ and $\tau_2 = (-x_o \sin \beta \sin \alpha + y_o \sin \beta \cos \alpha + z_o \cos \beta )|\mathbf{c}_o|^{-2}$.

Inserting \eqref{eq:P_Param} in \eqref{eq:orthogoCondParax}, we obtain the following expression for ensuring the orthogonality:
\begin{align}\label{eq:orthoParaxCond_1}
    &\sum_{m_1 = 1}^{M_1 } \sum_{m_2 = 1}^{M_2}
    \exp \biggl\{- j \frac{k_0}{|\mathbf{c}_o|} \biggl\{\left[ \tau_{11} \delta_1^t (u_1 - v_1) + \tau_{12} \delta_2^t (u_2 - v_2) \right] \nonumber\\
    &\ \; \times \delta_1^r \left(m_1 - \frac{M_1-1}{2}\right)  +
     \left[\tau_{21} \delta_1^t (u_1 - v_1) + \tau_{22} \delta_2^t (u_2 - v_2)\right] \nonumber\\
     & \ \; \times \delta_2^r \left(m_2 - \frac{M_2-1}{2}\right) \biggr\}\biggr\}=0 \qquad \forall (u_1, u_2) \not= (v_1, v_2) \, . 
\end{align}

The expression in~\eqref{eq:orthoParaxCond_1} can be simplified by using the geometric sum formula in~\cite[Eq. 3.1.10]{Abramowitz1965}, which results in the following orthogonality condition for all $(u_1, u_2) \not= (v_1, v_2)$:
\begin{align}
\label{eq:orthoParaxCond}
    &\frac{\sin \left[ \pi (\gamma_{11}(u_1 - v_1) + \gamma_{12}(u_2 - v_2)) \right]}
    {\sin \left[(\pi/M_1) (\gamma_{11}(u_1 - v_1) + \gamma_{12} (u_2 - v_2)) \right]} \nonumber \\
    & \qquad \times
   \frac{\sin \left[ \pi (\gamma_{21}(u_1 - v_1) + \gamma_{22}(u_2 - v_2)) \right]}
    {\sin \left[(\pi/M_2) (\gamma_{21}(u_1 - v_1) + \gamma_{22}(u_2 - v_2) ) \right]}=0  
\end{align}
where
\begin{equation}
    \gamma_{ab} = \frac{\tau_{ab} M_b}{\lambda |\mathbf{c}_o|} \delta_b^r \delta_a^{t}  \quad \forall a,b = \{1,2\} \, .
\end{equation}

In agreement with \cite{Boehagen2007}, we evince that the orthogonality condition presented in \eqref{eq:orthoParaxCond} admits a solution provided that at least one $\gamma_{ab}$ is equal to zero. This implies that it is not necessary to ensure the channel separability, i.e., to express $\mathbf{H}$ as the product of the channel matrices for each axis, in order to achieve the desired  orthogonality condition.

\subsection{Novelty and Insights from the Orthogonality Condition in  \texorpdfstring{\eqref{eq:orthoParaxCond}}{(25)}}

Compared with the orthogonality condition obtained in \cite{Boehagen2007}, which is expressed in terms of projections of the antenna arrays onto the planes that are perpendicular to the axis connecting their center-points, the condition in \eqref{eq:orthoParaxCond} is expressed in terms of parameters that are independent of the system of coordinates. Therefore, the obtained expression is simpler to be interpreted and to be used. If, for example, we wish to analyze the impact of translating the antenna array at the receiver by using the approach proposed in~\cite{Boehagen2007}, we would need to recompute the projections of both antenna arrays onto the plane perpendicular to the new system of coordinates. With the proposed formulation in \eqref{eq:orthoParaxCond}, this is not needed and the impact of a translation between the antenna arrays is apparent from the considered parametrization.

In addition, it is easier to identify the setups of parameters for which the orthgonality is achieved. Specifically, a close inspection of \eqref{eq:tau_2} reveals that the orthogonality condition in \eqref{eq:orthoParaxCond} can be formulated explicitly when either $\tau_{12} = 0$ or $\tau_{21} = 0$. This is elaborated next. It is pertinent to note that the conditions $\tau_{11} = 0$ or $\tau_{22} = 0$ lead to network deployments that either are of less practical relevance or are equivalent to the network deployments corresponding to $\tau_{12} = 0$ or $\tau_{21} = 0$. Therefore, these case studies are not discussed in this paper.

To obtain simple orthogonality criteria, let us further analyze \eqref{eq:orthoParaxCond}. Let us assume that there is a $\gamma_{ab} = 0$ for $a\not=b$. The function $\sin(\pi\gamma_{aa}(u_a-v_a))/\sin(\pi\gamma_{aa}(u_a-v_a)/M_a)$ is a periodic function with period $M_a/\gamma_{aa}$ and it has a zero in $n/\gamma_{aa}$ for $n \in (1, M_a-1)$. Then, we can ensure that the condition in \eqref{eq:orthogoCondParax} is fulfilled, for $(u_1 - v_1) \not= 0$, when $|\gamma_{aa}| = 1$ and $M_a\geq L_a$. When $(u_1 - v_1) = 0$, on the other hand, the orthogonality condition in \eqref{eq:orthogoCondParax} is fulfilled by configuring the array such that $|\gamma_{bb}| = 1$ and $M_b\geq L_b$. 
In summary, if the arrays are in a setup in which the orthogonality is possible, i.e., at least one $\gamma_{ab}$ can be set equal to zero, the following conditions need to hold simultaneously for ensuring that the channel orthogonality along the first and second principal directions of the antenna arrays is fulfilled:
\begin{align}
\label{eq:orthoCondParax0}
         \delta_1^r \delta_1^t &= \frac{\lambda |\mathbf{c}_o|}{M_1|\tau_{11}|} \:, \qquad &M_1 &\geq L_1 \\
        \delta_2^r \delta_2^t &= \frac{\lambda |\mathbf{c}_o|}{M_2|\tau_{22}|} \:, \qquad &M_2 &\geq L_2 \; .
\end{align}

It is worth mentioning that the orthogonality condition in \eqref{eq:orthoCondParax0} can be ensured by setting $|\gamma_{11}| = n_1$ and $|\gamma_{22}| = n_2$ for any positive integer $n_1$ and $n_2$. However, this results in large antenna arrays for which the paraxial approximation may not hold anymore. These setups are, therefore, of less interest from a practical point of view.

In the following, to gain further insights onto the explicit orthogonality conditions in \eqref{eq:orthoCondParax0}, we analyze the case studies in which the antenna arrays at the transmitter and receiver are aligned either along the $z$-axis or along the $x$-axis.

\subsubsection{Alignment along the \texorpdfstring{$z$-axis}{z-axis}} 
Let us assume that the antenna arrays at the transmitter and receiver are aligned along the $z$-axis, i.e., $z_o = 0$. Then, $\tau_{ab}$ can be simplified as follows:
\begin{align}
\tau_{11} &= \cos \alpha - x_o \frac{x_o \cos \alpha + y_o \sin \alpha}{|\mathbf{c}_o|^2}\\
 \tau_{12} &= 0 \\
\tau_{21} &= - \sin \beta \sin \alpha - x_o \frac{(-x_o \sin \beta \sin \alpha + y_o \sin \beta \cos \alpha)}{|\mathbf{c}_o|^2}\\
 \tau_{22} &= \cos \beta
\end{align}

Based on the obtained conditions, we note that $\tau_{12} = 0$. This implies that the matrix $\mathbf{H}$ can always be made orthogonal in this deployment. The inter-distances ensuring the orthogonality condition are those obtained by inserting the obtained $\tau_{11}$ and $\tau_{22}$ into \eqref{eq:orthoCondParax0}. Notably, we observe that $\tau_{11}$ and $\tau_{22}$ attain their maximum values when the two arrays are deployed in broadside, and hence, in this setup, the size of the antenna arrays is the smallest according to \eqref{eq:orthoCondParax0}.

\subsubsection{Alignment along the \texorpdfstring{$x$-axis}{x-axis}} 
Let us assume that the antenna arrays at the transmitter and receiver are aligned along the $x$-axis, i.e., $x_o = 0$. Then, $\tau_{ab}$ can be simplified as follows:
\begin{align} 
\tau_{11} &= \cos \alpha \\ 
\tau_{12} &= - z_o y_o \sin \alpha|\mathbf{c}_o|^{-2} \\
\tau_{21} &= -\sin \beta \sin \alpha \\ 
\tau_{22} &= \cos \beta - z_o (y_o \sin \beta \cos \alpha + z_o \cos \beta )|\mathbf{c}_o|^{-2}.
\end{align}

In this deployment scenario, none of the obtained $\tau_{ab}$ (or equivalently $\gamma_{ab}$) is equal to zero regardless of the considered system parameters. This asymmetry between the case studies in which the antenna arrays are aligned along the $z$-axis and $x$-axis is only due to the considered parametrization, which is formulated by first applying the tilt $\beta$ and then the rotation $\alpha$. By inverting these operations, the conditions obtained when the antenna arrays are aligned along the $z$-axis and the $x$-axis are swapped. Based on the obtained expressions for $\tau_{ab}$, the orthogonality conditions can be ensured, for example, by setting  $\alpha = 0$, i.e., $\tau_{12} = 0$ and $\tau_{21} = 0$. 

In summary, the main contributions of this section can be summarized as follows:
\begin{itemize}
    \item We have obtained explicit expressions for the optimal design of the antenna arrays in LoS MIMO channels, assuming the paraxial setting. The expressions are given in \eqref{eq:orthoCondParax0}, and they need to be fulfilled simultaneously.
    \item We have analyzed two case studies and have provided closed-form expressions for the inter-distances among the antenna elements of the arrays in order to ensure that the LoS MIMO channel has full rank.
    \item The obtained orthogonality conditions are formulated in an explicit manner and are simpler to interpret, for several network deployments of interest, as compared with the analysis carried out in \cite{Boehagen2007,Do2023}. This follows by comparing \eqref{eq:orthoParaxCond} with \cite[Eq. (24)]{Boehagen2007}, since the parameters $\gamma_{ab}$ in \eqref{eq:orthoParaxCond} are formulated explicitly as a function of the system parameters.
\end{itemize}

\section{Non-Paraxial Setting}\label{sec:NonParax}
Direct inspection of the orthogonality conditions in \eqref{eq:orthoCondParax0} shows that the paraxial approximation in \eqref{eq:paraxAssum} may not always be fulfilled. If, for example, the inter-distances at the multi-antenna transmitter are  $\delta^t_1 = \delta^t_2 = \lambda/2$, as in a typical user equipment, the inter-distances at the multi-antenna receiver need to be very large for ensuring that the orthogonality condition is fulfilled, assuming that \eqref{eq:orthoCondParax0} is still valid for the resulting LoS MIMO channel. Thus, the non-paraxial setting is a relevant case study, especially if one of the two antenna arrays has a compact size.

In this section, we identify the conditions to make the channel matrix $\mathbf{H}$ orthogonal, i.e., to fulfill the orthogonality condition in \eqref{eq:orthogoCondGen} in a non-paraxial setting. First, we introduce the channel model in the non-paraxial setting, and then we analyze the orthogonality condition in \eqref{eq:orthogoCondGen}.

\subsection{Non-Paraxial Channel Model}\label{subsec:NonParaxChannel}
When the paraxial condition in \eqref{eq:paraxAssum} is not fulfilled, we capitalize on the approach introduced in Section \ref{sec:SM}, which combines the quartic approximation with the sub-array partitioning method. 
Specifically, the large antenna array at the receiver is partitioned into $N^r$ sub-arrays. The $i$-th sub-array is centered in $\mathbf{c}^{r,i} = (x_o^i, y_o^i, z_o^i)$ and it has $M^i = M_1^i M_2^i$ antenna elements, where $M_1^i$ denotes the number of antenna elements in the first principal direction and $M_2^i$ denotes the numbers of antenna elements in the second principal direction. The position of the $m^i$-th antenna element is denoted by $\mathbf{r}^{r,i}_{m^i} = (x_{m^i}^{r,i},y_{m^i}^{r,i},z_{m^i}^{r,i}) + \mathbf{c}^{r,i}$. Based on the partition in sub-arrays, the channel matrix $\mathbf{H}$ can be rewritten as follows:
\begin{equation}
\label{eq:subArrayModel}
    \mathbf{H} = \begin{bmatrix} \mathbf{H}^{1} &
    \mathbf{H}^{2}&
    \hdots &
\mathbf{H}^{N^r}\end{bmatrix}^{{T}} 
\end{equation}
where $\mathbf{H}^{i} \in \mathbb{C}^{M^i \times L}$ is the channel matrix between the antenna array at the transmitter and the $i$-th antenna sub-array at the receiver. The size of the sub-arrays is chosen to ensure that the paraxial approximation can be applied to each sub-array, which implies the following: 
\begin{equation}
\label{eq:paraxAssumtion_sub}
    x^t_l, y^t_l, z^t_l, x_{m^i}^{r,i},y_{m^i}^{r,i},z_{m^i}^{r,i} \ll |\mathbf{c}_o^i| 
\end{equation}
where $|\mathbf{c}_o^i| = |\mathbf{c}^{r,i} - \mathbf{c}^t|$. For clarity, the channel matrix $\mathbf{H}$ in \eqref{eq:subArrayModel} is denoted by $\mathbf{H}^{\mathrm{Large}}$.

Accordingly, the quartic approximation can be applied to each sub-array, by considering a single system of coordinates for all the sub-arrays. Based on \eqref{eq:quarticApprox}, and by using the same line of thought as for \eqref{eq:GParax2}, the quartic approximation for $\mathbf{H}^{i}$ can be formulated as follows: 
\begin{equation}
\label{eq:subArrayChannel}
    \mathbf{H}^i \approx \frac{1}{4\pi|\mathbf{c}_o^i|} \mathbf{F}_{\mathrm{RX}}^i \mathbf{P}^i \left[ \mathbf{F}_{\mathrm{TX}}^i\right]^{*} 
\end{equation}
where $\mathbf{F}_{\mathrm{TX}}^i \in \mathbb{C}^{L \times L}$, $\mathbf{F}_{\mathrm{RX}}^i \in \mathbb{C}^{M^i \times M^i}$, and $\mathbf{P}^i \in \mathbb{C}^{M^i \times L}$ are defined as follows:
\mathleft
\begin{align}
\label{eq:focFunction_sub}
    \mathbf{F}_{\mathrm{TX}}^i(l,l) = \exp \Bigg\{ & j \frac{k_0}{2|\mathbf{c}_o^i|}\left[ (x^t_l)^2 + (y^t_l)^2 + (z^t_l)^2   - 2 x_o^i x^t_l   \vphantom{\frac{(z^t_l)^2}{|\mathbf{c}_o^i|^2}}\right. \nonumber\\ 
    & \left.   - 2 y_o^i y^t_l - 2 z_o^i z^t_l - \frac{(x_o^i x^t_l + z_o^i z^t_l)^2}{|\mathbf{c}_o^i|^2} \right]\Bigg\}
\end{align}
\begin{align}
    \mathbf{F}_{\mathrm{RX}}^i(m^i,m^i) = \exp \Bigg\{ & j \frac{k_0}{2|\mathbf{c}_o^i|}\left[ (x^{r,i}_{m^i})^2 + (y^{r,i}_{m^i})^2 + (z^{r,i}_{m^i})^2  
    \vphantom{\frac{(z^t_l)^2}{|\mathbf{c}_o^i|^2}}\right. \nonumber\\ 
    & \hspace{-2.5cm}\left.  + 2 x_o^i x^{r,i}_{m^i} + 2 y_o^i y^{r,i}_{m^i} + 2 z_o^i z^{r,i}_{m^i} - \frac{(x_o^i x^{r,i}_{m^i} + z_o^i z^{r,i}_{m^i})^2}{|\mathbf{c}_o^i|^2}\right]\Bigg\}
\end{align}
\begin{align}
    \mathbf{P}^i(m^i,l) = \exp \Bigg\{& -j \frac{k_0}{|\mathbf{c}_o^i|} \left[ x^{r,i}_{m^i} x^t_l + y^{r,i}_{m^i} y^t_l + z^{r,i}_{m^i} z^t_l 
     \vphantom{\frac{(z^t_l)^2}{|\mathbf{c}_o^i|^2}} \right.  \nonumber\\
    & \hspace{-1.8cm} \left. - \frac{(x_o^i x^{r,i}_{m^i} + y_o^i y^{r,i}_{m^i} + z_o^i z^{r,i}_{m^i} ) (x_o^i x^t_l + y_o^i y^t_l + z_o^i z^t_l) }{|\mathbf{c}_o^i|^2}\right]\Bigg\}
\end{align}
\mathcenter
and the off-diagonal elements of $\mathbf{F}_{\mathrm{TX}}^i$ and $\mathbf{F}_{\mathrm{RX}}^i$ are equal to zero, i.e., they are diagonal matrices.

\subsection{Channel Orthogonality}\label{subsec:NonOrthoParax}
To identify the orthogonality conditions in the non-paraxial setting, we need to analyze the matrix $\mathbf{G}^{\mathrm{Large}} = \left(\mathbf{H}^{\mathrm{Large}}\right)^* \mathbf{H}^{\mathrm{Large}}$ and to impose the equality in \eqref{eq:orthogoCondGen}. By inserting~\eqref{eq:subArrayModel} in \eqref{eq:orthogoCondGen}, we obtain the following:
\begin{equation}
\label{eq:orthoCondNP}
    \sum_{i = 1}^{N^r} \sum_{m^i=1}^{M^i} \left[\mathbf{H}^{i}(m^i,u)\right]^* \mathbf{H}^{i }(m^i, v) = 0 \quad \forall u \not= v \, .
\end{equation}

Furthermore, by inserting \eqref{eq:subArrayChannel} in \eqref{eq:orthoCondNP}, we obtain
\begin{align}
\label{eq:orthoCondNP2}
    \sum_{i = 1}^{N^r} & \sum_{m^i=1}^{M^i} \left[\frac{1}{4\pi|\mathbf{c}_o^i|} \mathbf{F}_{\mathrm{RX}}^i( m^i,m^i) \mathbf{P}^{i}(m^i,u)  \left[ \mathbf{F}_{\mathrm{TX}}^i(u,u) \right]^* \right]^*  \nonumber\\
    & \hspace{-0.25cm} \times \left[\frac{1}{4\pi|\mathbf{c}_o^i|} \mathbf{F}_{\mathrm{RX}}^i( m^i,m^i) \mathbf{P}^{i}(m^i,v)  \left[ \mathbf{F}_{\mathrm{TX}}^i(v,v) \right]^*\right] = 0 \quad \forall u \not= v \, .
\end{align}

Since $\left[\mathbf{F}_{\mathrm{RX}}^i( m^i,m^i) \right]^* \mathbf{F}_{\mathrm{RX}}^i( m^i,m^i) = 1$ by definition, \eqref{eq:orthoCondNP} simplifies to
\begin{align}
\label{eq:orthoCondNP3}
    \sum_{i = 1}^{N^r} &\frac{1}{|\mathbf{c}_o^i|^2} \mathbf{F}_{\mathrm{TX}}^i(u,u) \left[ \mathbf{F}_{\mathrm{TX}}^i(v,v) \right]^* \nonumber\\
    &\times \sum_{m^i=1}^{M^i} \left[ \mathbf{P}^{i}(m^i,u) \right]^*
    \left[\mathbf{P}^{i}(m^i,v)  \right] = 0 \quad \forall u \not= v \, .
\end{align}

In order to identify simple design criteria for ensuring the orthogonality of the LoS MIMO channel matrix, \eqref{eq:orthoCondNP3} needs to be simplified. To this end, we introduce the diagonal matrix $\bar{\mathbf{F}}_{\mathrm{Tx}} \in \mathbb{C}^{L\times L}$, whose diagonal entries are defined as follows:
\begin{align}
    \bar{\mathbf{F}}_{\mathrm{Tx}}(u,u) = \exp \Biggl\{  j \frac{k_o}{2|\mathbf{c}_o|}&\left[
(x^t_u)^2 + (y^t_u)^2 + (z^t_u)^2   \vphantom{\frac{(z^t_u)^2}{|\mathbf{c}_o|^2}} \right. \nonumber\\
& \hspace{1cm}\left.- \frac{(x_o x^t_u + z_o z^t_u)^2}{|\mathbf{c}_o|^2}\right]\Biggl\} \, .
\end{align}

Since, by definition, $\bar{\mathbf{F}}_{\mathrm{Tx}} \left[ \bar{\mathbf{F}}_{\mathrm{Tx}} \right]^* = \mathbf{I}_L$, \eqref{eq:orthoCondNP3} can be rewritten as follows:
\begin{align}
\label{eq:orthoCondNP3bis}
    \sum_{i = 1}^{N^r} \frac{1}{|\mathbf{c}_o^i|^2} &\left[\mathbf{F}_{\mathrm{TX}}^i(u,u) \left[\bar{\mathbf{F}}_{\mathrm{Tx}}(u,u)\right]^*\right] \left[ \mathbf{F}_{\mathrm{TX}}^i(v,v) \left[\bar{\mathbf{F}}_{\mathrm{Tx}}(v,v)\right]^*\right]^* \nonumber \\
    &\times \sum_{m^i=1}^{M^i} \left[ \mathbf{P}^{i}(m^i,u) \right]^*
    \left[\mathbf{P}^{i}(m^i,v)  \right] = 0 \quad \forall u \not= v \, .
\end{align}

Let us analyze the product $\mathbf{F}_{\mathrm{TX}}^i(u,u) \left[\bar{\mathbf{F}}_{\mathrm{Tx}}(u,u)\right]^* = \exp \left\{ {j{k_o}\Phi _{{\rm{TX}}}^i\left( {u,u} \right)} \right\}$, where $\Phi _{{\rm{TX}}}^i\left( {u,u} \right)$ is defined as follows:
\begin{align} 
\Phi _{{\rm{TX}}}^i\left( {u,u} \right) &= \Psi _{{\rm{TX}}}^i\left( {u,u} \right) + \Delta _{{\rm{TX}}}^i\left( {u,u} \right) -  \Delta _{{\rm{TX}}}^o\left( {u,u} \right) \\ 
&\approx \Psi _{{\rm{TX}}}^i\left( {u,u} \right) \label{eq:GeneralApproximation_1}
\end{align}
with 
\begin{align} \label{eq:GeneralApproximation_2}
& \Psi _{{\rm{TX}}}^i\left( {u,u} \right) = -\frac{1}{2|\mathbf{c}_o^i|}\left[ x_o^i x^t_u + y_o^i y^t_u + z_o^i z^t_u \right] \\
& \Delta _{{\rm{TX}}}^i\left( {u,u} \right) = \frac{1}{2|\mathbf{c}_o^i|}\left[ (x^t_u)^2 + (y^t_u)^2 + (z^t_u)^2 - \frac{(x_o^i x^t_u + z_o^i z^t_u)^2}{|\mathbf{c}_o^i|^2} \right] \\
& \Delta _{{\rm{TX}}}^o\left( {u,u} \right) = \frac{1}{2|\mathbf{c}_o|}\left[ (x^t_u)^2 + (y^t_u)^2 + (z^t_u)^2 - \frac{(x_o x^t_u + z_o z^t_u)^2}{|\mathbf{c}_o|^2} \right] \, .
\end{align}

The rationale for the approximation in \eqref{eq:GeneralApproximation_1} is the following:
\begin{itemize}
\item \textit{Sub-arrays located around the center-point of the multi-antenna receiver}. In this case, it holds $|\mathbf{c}_o^i| \approx |\mathbf{c}_o|$. As a result, $\Delta _{{\rm{TX}}}^i\left( {u,u} \right) \approx \Delta _{{\rm{TX}}}^o\left( {u,u} \right)$, and, therefore, $\Psi _{{\rm{TX}}}^i\left( {u,u} \right) \gg \Delta _{{\rm{TX}}}^i\left( {u,u} \right) -  \Delta _{{\rm{TX}}}^o\left( {u,u} \right)$. Equation \eqref{eq:GeneralApproximation_1} is then a good approximation in this case.
\item \textit{Sub-arrays located far away from the center-point of the multi-antenna receiver}. In this case, it holds $|\mathbf{c}_o^i| \gg |\mathbf{c}_o|$. In $\Psi _{{\rm{TX}}}^i\left( {u,u} \right)$, the values of $x_o^i$, $y_o^i$ and $z_o^i$ scale similar to $|\mathbf{c}_o^i|$, as the latter increases. Therefore, $\Psi _{{\rm{TX}}}^i\left( {u,u} \right)$ does not become arbitrarily small as $|\mathbf{c}_o^i|$ increases. In $\Delta _{{\rm{TX}}}^i\left( {u,u} \right)$ and $\Delta _{{\rm{TX}}}^o\left( {u,u} \right)$, $x_u^t$, $y_u^t$ and $z_u^t$ fulfill the paraxial approximation in \eqref{eq:paraxAssumtion_sub}, depend on the transmitter, and are independent of $|\mathbf{c}_o^i|$. As $|\mathbf{c}_o^i|$ increases, therefore, $\Delta _{{\rm{TX}}}^i\left( {u,u} \right)$ tends to be very small, i.e., $\Delta _{{\rm{TX}}}^i\left( {u,u} \right) \ll \Psi _{{\rm{TX}}}^i\left( {u,u} \right)$. As for $\Delta _{{\rm{TX}}}^o\left( {u,u} \right)$, we note that, e.g., $\left( {x_o^i/\left| {{\bf{c}}_o^i} \right|} \right)x_u^t$ in $\Psi _{{\rm{TX}}}^i\left( {u,u} \right)$ is much greater than ${\left( {x_u^t} \right)^2}/\left| {{\bf{c}}_o^i} \right| = \left( {x_u^t/\left| {{{\bf{c}}_o}} \right|} \right)x_u^t$ in $\Delta _{{\rm{TX}}}^o\left( {u,u} \right)$, since $x_o^i$ scales similar to $|\mathbf{c}_o^i|$ while $x_u^t/\left| {{{\bf{c}}_o}} \right| \ll 1$ according to the paraxial approximation in \eqref{eq:paraxAssumtion_sub}. Similar inequalities can be applied to the other addends. Therefore, $\Delta _{{\rm{TX}}}^o\left( {u,u} \right) \ll \Psi _{{\rm{TX}}}^i\left( {u,u} \right)$. Equation \eqref{eq:GeneralApproximation_1} is then a good approximation in this case.
\end{itemize}

Based on the approximation in \eqref{eq:GeneralApproximation_1}, the orthogonality condition in \eqref{eq:orthoCondNP3bis} can be expressed as
\begin{align}
\label{eq:orthoCondNP3_bis}
    \sum_{i = 1}^{N^r}& \frac{1}{|\mathbf{c}_o^i|^2} \exp \Bigg\{ -j \frac{k_0}{|\mathbf{c}_o^i|}\left[ x_o^i (x^t_u - x^t_v)   \right. \nonumber\\
    & \hspace{2cm}\left. + y_o^i (y^t_u - y^t_v) + z_o^i (z^t_u - z^t_v)\right]\Bigg\}  \nonumber\\
    &\hspace{0.25cm} \times \sum_{m^i=1}^{M^i} \left[ \mathbf{P}^{i}(m^i,u) \right]^*
    \left[\mathbf{P}^{i}(m^i,v)  \right] = 0 \quad \forall u \not= v \, .
\end{align} 

By comparing the obtained orthogonality condition in \eqref{eq:orthoCondNP3_bis} for the non-paraxial setting with the akin one in \eqref{eq:orthogoCondParax} for the paraxial setting, we identify a major difference: The orthogonality condition in \eqref{eq:orthoCondNP3} depends on an exponential term that originates from the matrix $\mathbf{F}_{\mathrm{TX}}^i$, as a result of the partitioning in sub-arrays at the multi-antenna receiver. This makes the optimal designs to realize a full-rank LoS MIMO channel in the paraxial and non-paraxial settings different. In the paraxial setting, specifically, we have proved that uniform antenna arrays, in which the inter-distances $\delta^r_1$ and $\delta^r_2$ are the same between all the antenna elements, are sufficient for ensuring that the matrix $\mathbf{H}$ is orthogonal. In the non-paraxial setting, on the other hand, the inter-distances in each sub-array are expected to be different because of the different distances between the antenna array at the transmitter and each antenna sub-array at the receiver. For generality, the inter-distances among the antenna elements of the $i$-th sub-array are denoted by $\delta_1^{r, i}$ and $\delta_2^{r, i}$ along the first and second principal directions of the antenna array at the receiver, respectively.

Accordingly, we introduce the following parametrization for the position of the $m^i$-th antenna element in the $i$-th sub-array at the receiver:
\begin{align}
    \mathbf{r}_{m^i}^{r, i} = &\left( \delta_1^{r, i} m_1^{c,i} \cos \alpha - \delta_2^{r, i} m_2^{c,i} \sin \beta \sin \alpha, \delta_1^{r, i} m_1^{c,i} \sin \alpha\right. \nonumber \\
    & \hspace{0.25cm}\left.  + \delta_2^{r, i} m_2^{c,i} \sin \beta \cos \alpha,
    \delta_2^{r, i} m_2^{c,i} \cos\beta \right) + \mathbf{c}^{r,i}
\end{align}
where  $m_1^{c,i} = m_1^i - (M_1^i-1)/2$, $m_2^{c,i} = m_2^i - (M_2^i-1)/2$ and $m^i = (m_1^i - 1) M_1^i + m_2^i$, with $m_1^i$ and $m_2^i$ denoting the indices of the $i$-th sub-array along the first and second principal directions.

The inner summation in \eqref{eq:orthoCondNP3} can be computed by utilizing the same approach as in Section \ref{subsec:OrthoParax}, which results in the following analytical expression:
\begin{align}
    \sum_{m^i=1}^{M^i}& \left[ \mathbf{P}^{i}(m^i,u) \right]^* \left[\mathbf{P}^{i}(m^i,v)  \right] \nonumber
\end{align}
\begin{align}
\label{eq:productP}
&=\frac{\sin \left[ \pi (\gamma_{11}^{i}(u_1 - v_1) + \gamma_{12}^{i}(u_2 - v_2)) \right]}
    {\sin \left[(\pi/M_1^{i}) (\gamma_{11}^{i}(u_1 - v_1) + \gamma_{12}^{i} (u_2 - v_2)) \right]} \nonumber \\
    &\quad \times \frac{\sin \left[ \pi (\gamma_{21}^{i}(u_1 - v_1) + \gamma_{22}^{i}(u_2 - v_2)) \right]}
    {\sin \left[(\pi/M_2^{i}) (\gamma_{21}^{i}(u_1 - v_1) + \gamma_{22}^{i}(u_2 - v_2) ) \right]}
\end{align}
where
\begin{equation}
\label{eq:gamma_sub}
    \gamma_{ab}^{i} = \frac{\tau_{ab}^{i} M_b^i}{\lambda |\mathbf{c}_o^i|} \delta_b^{r,i} \delta_a^{t}  \quad \forall a,b = \{1,2\} \, .
\end{equation}
Inserting {\eqref{eq:paramTx} and} \eqref{eq:productP} into \eqref{eq:orthoCondNP3_bis}, we obtain
%
\begin{align}
\label{eq:orthoCondNP4}
    \sum_{i = 1}^{N^r} &\frac{1}{|\mathbf{c}_o^i|^2} \exp \Big\{ -j \pi \left[ \eta_x^i(u_1 - v_1) + \eta_z^i (u_2 - v_2)\right]\Big\} \nonumber\\
    &\times  \frac{\sin \left[ \pi (\gamma_{11}^{i}(u_1 - v_1) + \gamma_{12}^{i}(u_2 - v_2)) \right]} 
    {\sin \left[(\pi/M_1^{i}) (\gamma_{11}^{i}(u_1 - v_1) + \gamma_{12}^{i} (u_2 - v_2)) \right]} \nonumber \\
    & \times \frac{\sin \left[ \pi (\gamma_{21}^{i}(u_1 - v_1) + \gamma_{22}^{i}(u_2 - v_2)) \right]}
    {\sin \left[(\pi/M_2^{i}) (\gamma_{21}^{i}(u_1 - v_1) + \gamma_{22}^{i}(u_2 - v_2) ) \right]} = 0
\end{align}
for all $(u_1, u_2) \not= (v_1, v_2)$, with
\begin{equation}
\label{eq:eta_def}
    \eta_x^i = \frac{2 x_o^{r,i}}{\lambda |\mathbf{c}_o^i|} \delta_1^t \:, \qquad \eta_z^i = \frac{2 z_o^{r,i}}{\lambda |\mathbf{c}_o^i|} \delta_2^t \, .
\end{equation}

Equation \eqref{eq:orthoCondNP4} generalizes the orthogonality condition in \eqref{eq:orthoParaxCond} to non-paraxial settings. In general, the optimal inter-distances that maximize the rank of the channel matrix $\mathbf{H}$ can be obtained by solving \eqref{eq:orthoCondNP4} numerically. To obtain some design insights, we consider the case study of linear arrays with broadside orientation, i.e., the two antenna arrays are parallel to one another and are faced to each other.

\subsection{Explicit Orthogonality Conditions for Linear Arrays with Broadside Orientation} \label{sec:BroadsideLinear}
Let us consider the case study in which the antenna arrays at the transmitter and receiver are linear, i.e., $M^i_2 = L_2 = 1$, are parallel and are faced to each other (broadside orientation), i.e., $\alpha = 0$, $\beta = 0$ and $\mathbf{c}_o = (0, y_o, 0)$. Thus, we have $u_1 = u$ and $v_1 = v$, and \eqref{eq:orthoCondNP4} simplifies to
\begin{align}
\label{eq:orthoCondULAs}
    \sum_{i = 1}^{N^r} \frac{1}{|\mathbf{c}_o^i|^2} & \exp \Big\{ -j \pi \eta_x^i(u - v) \Big\}  \nonumber \\
    & \times\frac{\sin \left[ \pi \gamma_{11}^{i}(u - v) \right]} 
    {\sin \left[(\pi/M_1^{i}) \gamma_{11}^{i}(u - v) \right]}  = 0 \quad \forall u \not= v \, .
\end{align}

In addition, to enhance the analytical tractability and the design insights from it, we assume that the number of sub-arrays at the receiver is even, and that the centers of the sub-arrays are distributed symmetrically with respect to the $yz$-plane, i.e., $x^{r,i}_o = -x^{r, (N_r + 1 - i)}_o$. Due to the symmetry of the considered deployment, the inter-distance between the antenna elements and the number of antenna elements are the same in the $i$-th and the $(N_r + 1 - i)$-th sub-arrays i.e., $\gamma_{11}^{i} = \gamma_{11}^{N_r + 1 - i}$ and $M_1^{i} = M_1^{N_r + 1 - i}$, respectively. Then, \eqref{eq:orthoCondULAs} can be simplified to
\begin{align}
    \label{eq:orthoCondULAs2}
    \sum_{i = 1}^{N^r/2} \frac{1}{|\mathbf{c}_o^i|^2} & \cos \left[\pi |\eta_x^i| (u - v)\right] \nonumber \\ &\times
    \frac{\sin \left[ \pi \gamma_{11}^{i}(u - v) \right]} 
    {\sin \left[(\pi/M_1^{i}) \gamma_{11}^{i}(u - v) \right]}
     = 0 \quad \forall u \not= v \, .
\end{align}

It is worth mentioning that a sufficient condition for fulfilling \eqref{eq:orthoCondULAs2} consists of independently optimizing the arrangements of the sub-arrays such that the condition
\begin{equation}
\frac{\sin \left[ \pi \gamma_{11}^{i}(u - v) \right]} {\sin \left[(\pi/M_1^{i}) \gamma_{11}^{i}(u - v) \right]} = 0
\qquad \forall i= 1,2,..., \frac{N^r}{2}
\end{equation}
is fulfilled for every sub-array at the receiver. However, similar to Section \ref{sec:Parax}, this condition can only be ensured if the number of antenna elements in each sub-array is at least equal to the number of antenna elements at the transmitter, i.e., $M^i_1 \geq L_1$, as well as if the condition $|\gamma_{11}^{i}| = 1$ $\forall i \in [1,N_r/2]$ is satisfied. For ensuring the channel orthogonality condition and a full-rank channel matrix, therefore, this simple solution requires a number of antenna elements at the receiver that is much larger than the number of antenna elements at the transmitter. In addition, this simple approach may lead to solutions that do not satisfy the considered modeling assumptions, since  the large size of the resulting sub-arrays may not fulfill the paraxial approximation condition in \eqref{eq:paraxAssumtion_sub}, hence invalidating the application and accuracy of the proposed quartic approximation of the wavefront.

A more suitable design criterion, which reduces the number of antennas at the receiver for obtaining a full-rank LoS MIMO channel matrix, can be obtained by jointly optimizing the arrangements of all the sub-arrays accordingly to \eqref{eq:orthoCondULAs2}. To this end, we rewrite \eqref{eq:orthoCondULAs2} by using the identity $\cos(\alpha) \sin(\beta) = (\sin(\beta + \alpha) + \sin(\beta - \alpha))/2$, as follows:
\begin{equation}
\label{eq:orthoCondULAs3}
    \sum_{i = 1}^{N^r/2} f^{i}_+(u-v) + f^{i}_-(u-v) = 0 \quad \forall u \not= v
\end{equation}
where
\begin{equation}
    f^{i}_+(u-v) =  \frac{1}{|\mathbf{c}_o^i|^2}\frac{\left[\sin \left[ \pi (\gamma_{11}^{i} + |\eta_x^i|)(u - v) \right]\right]} 
    {\sin \left[(\pi/M_1^{i}) \gamma_{11}^{i}(u - v) \right]}
\end{equation}
\begin{equation}
    f^{i}_-(u-v) =  \frac{1}{|\mathbf{c}_o^i|^2}\frac{\left[\sin \left[ \pi (\gamma_{11}^{i} - |\eta_x^i|)(u - v) \right]\right]} 
    {\sin \left[(\pi/M_1^{i}) \gamma_{11}^{i}(u - v) \right]} \, .
\end{equation}

Next, we demonstrate how \eqref{eq:orthoCondULAs3} can be utilized for optimizing the center-points of the sub-arrays and the inter-distances of the antenna elements in each sub-array. Also, we estimate the minimum number of antennas at the receiver for ensuring that the LoS MIMO channel matrix has a full rank. For ease of presentation, we progressively examine the cases studies for two, four and $N^r$ sub-arrays.

\subsubsection{Two sub-arrays}
Let us assume $N^r = 2$. Then, \eqref{eq:orthoCondULAs3} can be simplified as follows:
\begin{align}
    \label{eq:orthoCondULAs_2s}
      f^{1}_+(u-v) + f^{1}_-(u-v)  = 0 \quad \forall u \not= v \, .
\end{align}

A sufficient condition to satisfy \eqref{eq:orthoCondULAs_2s} consists of imposing the following two conditions:
\begin{align}
&f^{i}_-(u-v) = 0 \quad {\rm{for}} \; (u - v) = 1,2,...,L_1-1\\
&f^{i}_+(u-v) = 0 \quad {\rm{for}} \; (u - v) = 1,2,...,L_1-1
\end{align}
which leads to the following sufficient orthogonality conditions: 
\begin{align}
\label{eq:orthoCondULAs_2s_a}
 (\gamma_{11}^{1} - |\eta_x^1|) = 0, \qquad  (\gamma_{11}^{1} + |\eta_x^1|) = 1 
\end{align}
provided that $M^1_1/\gamma_{11}^{1} \geq L_1$.

By solving the systems of two equations in \eqref{eq:orthoCondULAs_2s_a}, we obtain the following conditions:
\begin{equation}
\label{eq:2ArrayCond}
\gamma_{11}^{1} = \frac{1}{2}, \qquad \qquad
    |\eta_x^1| = \frac{1}{2},  \qquad \qquad
    2M^1_1 \geq L_1  \, . 
\end{equation}

If $N^r = 2$, the only possible sub-array partitioning is $M^1_1 = M_1/2$. From \eqref{eq:2ArrayCond}, we obtain $2M^1_1 = M_1 \geq L_1$. This implies that it is sufficient that the number of antenna elements at the receiver is at least equal to the number of antenna elements at the transmitter for ensuring a full-rank LoS MIMO channel matrix when $N^r = 2$. 

By considering that $|\mathbf{c}^{1}|^2 = y_o^2 + (x^{r,1}_o)^2$ and by inserting the expressions for $\gamma_{11}^{1}$ and $|\eta_x^1|$ in \eqref{eq:2ArrayCond} into \eqref{eq:gamma_sub} and \eqref{eq:eta_def}, we obtain the following closed-form and explicit expressions for the center-points of the sub-arrays and for the inter-distances of the antenna elements:
\begin{align}
 |x_o^{r,1}| =  \frac{y_o}{\sqrt{(4\delta_1^{t}/\lambda)^2-1}}, \qquad \quad  \delta_1^{r,1} = \frac{{\lambda \sqrt{ y_o^2 + (x_o^{r,1})^2}}}{|{\tau}_{11}^{1}| M_1 \delta_1^{t}} \label{eq:2ArrayCond1}
\end{align}
provided that $M_1 \geq L_1$.

\subsubsection{Four sub-arrays} \label{subsubsec:NonParax_4s}
Let us assume $N^r = 4$. Then, \eqref{eq:orthoCondULAs3} can be simplified as follows:
\begin{align}
    \label{eq:orthoCondULAs_4s}
      &f^{1}_+(u-v) + f^{1}_-(u-v) \nonumber \\ 
      & \ + f^{2}_+(u-v) + f^{2}_-(u-v) = 0 \quad \forall u \not= v \, .
\end{align}

We utilize a similar approach as for the case study with two sub-arrays. Specifically, a sufficient condition to satisfy \eqref{eq:orthoCondULAs_4s} consists of imposing the following two conditions:
\begin{align}
    & f^{1}_-(u-v) + f^{2}_+(u-v) = 0 \quad \forall u \not= v \label{eq:orthoCondULAs_4s__1} \\
 & f^{1}_+(u-v) + f^{2}_-(u-v) = 0 \quad \forall u \not= v \, .   \label{eq:orthoCondULAs_4s__2}
\end{align}

The equality in \eqref{eq:orthoCondULAs_4s__1} can be approximately fulfilled, i.e., $f^{1}_-(u-v) \approx - f^{2}_+(u-v)$, if the following conditions are satisfied:
\begin{align}
   & (\gamma_{11}^{1} - |\eta_x^1|) = -(\gamma_{11}^{2} + |\eta_x^2|) \label{eq:4Array_3}\\
   & \frac{M_1^1}{|\mathbf{c}_o^1|^2} \frac{ (\gamma_{11}^{1} - |\eta_x^1|)}{\gamma_{11}^{1}} = - \frac{M_1^2}{|\mathbf{c}_o^2|^2} \frac{ (\gamma_{11}^{2} + |\eta_x^2|)}{\gamma_{11}^{2}} \label{eq:4Array_4} \\
& M_1^1 / \gamma_{11}^{1} > (L_1 -1), \quad M_1^2 / \gamma_{11}^{2} > (L_1 -1) \, .
\end{align}

In detail, \eqref{eq:4Array_3} is obtained by matching the zeros of the two functions $f^{1}_-(u-v)$ and $f^{2}_+(u-v)$, and \eqref{eq:4Array_4} is obtained by matching, with opposite signs, the peak amplitudes of $f^{1}_-(u-v)$ and $f^{2}_+(u-v)$. This approach based on matching the zeros and the peak amplitudes of $f^{1}_-(u-v)$ and $f^{2}_+(u-v)$ leads to the approximation $f^{1}_-(u-v) \approx - f^{2}_+(u-v)$.

In addition, a sufficient condition to satisfy the equality in \eqref{eq:orthoCondULAs_4s__2} can be obtained by applying the same approach as for the two sub-array case, which results in the following conditions:
\begin{align}
    \gamma_{11}^{1} + |\eta_x^1| = 1, \qquad \gamma_{11}^{2} - |\eta_x^2| = 0 \, . \label{eq:4Array_2}
\end{align}
provided that $M_1^1 / \gamma_{11}^{1} > (L_1 - 1)$.

By inserting \eqref{eq:4Array_2} in \eqref{eq:4Array_3} and \eqref{eq:4Array_4}, the set of equations that need to be satisfied for ensuring that the LoS MIMO channel matrix has full rank is the following:
\begin{align}
    &\gamma_{11}^{1} = 1 - |\eta_x^1|, \qquad
    \gamma_{11}^{2} = |\eta_x^2|, \qquad  |\eta_{x}^{2}| = |\eta_{x}^{1}| - \frac{1}{2} \label{eq:4Array_3b}\\
    &\frac{M_1^1}{|\mathbf{c}_o^1|^2} \frac{ 1 - 2|\eta_x^1|}{1 - |\eta_x^1|} = - \frac{2M_1^2}{|\mathbf{c}_o^2|^2}  \label{eq:4Array_4b}
\end{align}
provided that $M_1^i / \gamma_{11}^{i} > (L_1 -1)$, which is the minimum number of antenna elements in each sub-array for ensuring the orthogonality of the LoS MIMO channel.

The obtained system of equations can be solved by noting that $\gamma_{11}^{i}$ depends on $\delta_1^{r, i}$ and $|x^{r,i}_o|$, but $\eta_x^i$ depends only on $|x^{r,i}_o|$. Therefore, the solution that fulfills \eqref{eq:4Array_3b} and \eqref{eq:4Array_4b} can be obtained by first computing the center-points of the sub-arrays $|x^{r,1}_o|$ and $|x^{r,2}_o|$ from \eqref{eq:4Array_3b}, \eqref{eq:4Array_4b} and \eqref{eq:eta_def}, and then computing the corresponding inter-distances for each sub-array, $\delta_1^{r, 1}$ and $\delta_1^{r, 2}$, according to \eqref{eq:eta_def}. More specifically, the following identity can be obtained from \eqref{eq:eta_def}:
\begin{equation}
    \frac{y_o^2}{|\mathbf{c}_o^i|^2} = 1 - \left(\frac{|x_o^{r,i}|}{|\mathbf{c}_o^i|}\right)^2 = 1 - \left(\frac{|\eta_x^i| \lambda}{2\delta^t}\right)^2 \, .
\end{equation}

Considering this latter identity, \eqref{eq:4Array_4b} can be rewritten as follows:
\begin{equation}
\label{eq:4Array_4c}
        M_1^1 \left[ 1 - \left(\frac{|\eta_x^1| \lambda}{2\delta^t}\right)^2\right] \frac{ 1 - 2|\eta_x^1|}{1 - |\eta_x^1|} = - 2 M_1^2 \left[ 1 - \left(\frac{|\eta_x^2| \lambda}{2\delta^t}\right)^2\right]
\end{equation}
and \eqref{eq:4Array_4c} can be expressed only in terms of $|\eta_x^1|$ by inserting \eqref{eq:4Array_3b} in \eqref{eq:4Array_4c}, as follows:
\begin{align}
\label{eq:4Array_4d}
        &M_1^1 \left[ \left(\frac{2\delta^t}{ \lambda}\right)^2 - \left(|\eta_x^1|\right)^2\right] \frac{ 1 - 2|\eta_x^1|}{1 - |\eta_x^1|} \nonumber \\ 
        & \hspace{2.5cm}= - 2 M_1^2 \left[ \left(\frac{2\delta^t}{ \lambda}\right)^2 - \left(|\eta_x^1|-\frac{1}{2}\right)^2\right] \, .
\end{align}

Equation \eqref{eq:4Array_4d} is a cubic equation in terms of the unknown $|\eta_x^1|$, and it can hence be solved by using Cardano's formula~\cite[Eq. 3.8.2]{Abramowitz1965}. In detail, if $|\eta_x^1| \in (0,2\delta^t/\lambda)$, the only valid root of \eqref{eq:4Array_4d} needs to lie in the interval $(0,2\delta^t/\lambda)$. If there is no root in $(0,2\delta_1^t/\lambda)$, therefore, no optimal, i.e., full-rank, design for the considered array configuration exists. 

\textbf{Case study $\bm {\delta_1^t = \lambda} /\bm{2}$} -- In non-paraxial deployments, a critical case is constituted by the setting $\delta_1^t = \lambda/2$, which is the typical inter-distance in conventional antenna arrays. This is because the size of the antenna array at the receiver is assumed to be the largest one. In this case, \eqref{eq:4Array_4d} simplifies as follows:
\mathleft
\begin{equation}
        M_1^1 \left[ 1 - \left(|\eta_x^1|\right)^2\right] \frac{ 1 - 2|\eta_x^1|}{1 - |\eta_x^1|} = - 2 M_1^2 \left[ 1 - \left(|\eta_x^1|-\frac{1}{2}\right)^2\right]
\end{equation}
\mathcenter
and $|\eta_x^1| \in (0,2\delta^t/\lambda) = (0,1)$.

Discarding the root $|\eta_x^1| = 1$, as it is not in the feasible set, we obtain the following equation:
\begin{multline}
\label{eq:4Array_4e}
        \left(2M_1^1 + 2M_1^2\right)(|\eta_x^1|)^2 + \left(M_1^1 - 2M_1^2\right)|\eta_x^1| \\- M_1^1 - \frac{3}{2}M_1^2 = 0 \, .
\end{multline}

Given that $M_1 = 2 M_1^1 + 2 M_1^2$, the positive root of \eqref{eq:4Array_4e} is the following:
\begin{equation}
\label{eq:4subArraysEta}
    |\eta_x^1| = \frac{2M_1^2 - M_1^1}{2 {{M_1}}} + \frac{1}{2 {{M_1}}}\sqrt{9 (M_1^1)^2 + 16 M_1^1 M_1^2 + 16 (M_1^2)^2} \, .
\end{equation}

As mentioned, $|\eta_x^1|< 2\delta_1^t/\lambda = 1$ for being feasible. Therefore, we conclude that the sub-array partitioning needs to fulfill the following condition:
\begin{equation}
    3 M_1^2 < 4 M_1^1
\end{equation}
which is obtained from \eqref{eq:4subArraysEta} by imposing $|\eta_x^1|< 1$. The obtained expression highlights that some partitionings in sub-arrays are not feasible with the proposed approach.

\textbf{Minimum number of required antenna elements} -- In addition, the solution of the system of equations in  \eqref{eq:4Array_3b} and \eqref{eq:4Array_4b} needs to fulfill the condition $M_1^i > \gamma_{11}^i (L_1 -1)$, which imposes a minimum number of antenna elements in each sub-array. Based on \eqref{eq:4Array_3b}, the inequality $M_1^i > \gamma_{11}^i (L_1 -1)$ can be formulated in terms of $|\eta_x^1|$, as follows:
\begin{equation}
    M_1^1 > \left(1-|\eta_x^1|\right) \left(L_1-1\right), \  \:
    M_1^2 > \left(|\eta_x^1| - 1/2\right) \left(L_1-1\right) \label{eq:condSol_4s}
\end{equation}
where $|\eta_x^1|$ is given in \eqref{eq:4subArraysEta}. From \eqref{eq:4subArraysEta}, we obtain $|\eta_x^1| > 1/2$ for any $M_1^1$ and $M_1^2$. If $3 M_1^2 < 4 M_1^1$, therefore, we have $1/2 < |\eta_x^1| < 1$, and the right-hand sides of \eqref{eq:condSol_4s} are both positive. 

In conclusion, the proposed approach provides a LoS MIMO channel matrix with a full rank equal to $L_1$ if the numbers of antenna elements $M_1^1$ and $M_1^2$ fulfill the set of inequalities
\begin{align}
    M_1^1 & > \left(1 - \frac{2M_1^2 - M_1^1}{{{4M_1^1 + 4M_1^2}}} \right. \nonumber \\
    & \hspace{0.9cm}\left.- \frac{\sqrt{9 (M_1^1)^2 + 16 M_1^1 M_1^2 + 16 (M_1^2)^2}}{{{4M_1^1 + 4M_1^2}}}  \right) \left(L_1-1\right) \label{eq:FeasibilitySystem_1}  \\ 
    M_1^2 & > \left(\frac{2M_1^2 - M_1^1}{{{4M_1^1 + 4M_1^2}}} \right. \nonumber \\ 
    & \hspace{0.3cm} \left.+ \frac{\sqrt{9 (M_1^1)^2 + 16 M_1^1 M_1^2 + 16 (M_1^2)^2}}{{{4M_1^1 + 4M_1^2}}}  - \frac{1}{2}\right) \left(L_1-1\right) \label{eq:FeasibilitySystem_2}  \\
& \hspace{-0.4cm} 4 M_1^1 > 3 M_1^2  \label{eq:FeasibilitySystem_3} \\
& \hspace{-0.4cm} 2M_1^1 + 2M_1^2 \ge L_1 
\label{eq:FeasibilitySystem_4}
\end{align}
where the last inequality in \eqref{eq:FeasibilitySystem_4} ensures that the total number of antenna elements at the multi-antenna receiver is at least equal to the number of antenna elements at the multi-antenna transmitter, which is a necessary condition for obtaining a rank equal to $L_1$.

It is of particular interest to evaluate whether the design that requires the minimum number of antenna elements at the multi-antenna receiver is feasible, i.e., the MIMO configuration $2M_1^1 + 2M_1^2 = L_1$. In this case, $M_1^2 = L_1/2 - M_1^1$. By inserting the latter equality in \eqref{eq:FeasibilitySystem_3} and noting that $M_1^2 = L_1/2 - M_1^1 > 0$ by definition, we obtain $3 L_1/14 < M_1^1 < L_1/2$. By direct inspection of \eqref{eq:FeasibilitySystem_1} and \eqref{eq:FeasibilitySystem_2} with $M_1^2 = L_1/2 - M_1^1$, it is apparent that the two inequalities are not always fulfilled for any values of $L_1$, by assuming $3 L_1/14 < M_1^1 < L_1/2$. 

Therefore, the condition $2M_1^1 + 2M_1^2 = L_1$ needs to be relaxed with the inequality in \eqref{eq:FeasibilitySystem_4}. As an example, we illustrate a simple design criterion that is analyzed numerically in Section \ref{sec:NR}. Let us assume that the four sub-arrays have the same number of antenna elements, i.e., $M_1^1 = M_1^2 = M_1^0$. From $2M_1^1 + 2M_1^2 = M_1$, we then obtain $M_1^0 = M_1/4$. This design criterion has the positive feature that $|\eta_x^1|$ in \eqref{eq:4subArraysEta} is independent of $M_1^1$, $M_1^2$ and $M_1$. The computed value of $|\eta_x^1|$ can then be inserted into \eqref{eq:condSol_4s}, by obtaining $\bar M_1^1 = \left(1-|\eta_x^1|\right) \left(L_1-1\right)$, and $\bar M_1^2 = \left(|\eta_x^1| - 1/2\right) \left(L_1-1\right)$. Then, $M_1^0$ can be obtained as $M_1^0 > \max \left\{ {\bar M_1^1,\bar M_1^2} \right\}$, and $M_1 = 4 M_1^0$. In this case, $|\eta_x^1| = 0.925$, $M_1^0 > \bar M_1^2 = 0.4250(L_1-1)$, and $M_1 > 1.7(L_1-1)$. This case study is further illustrated in Section \ref{sec:NR} with the aid of numerical simulations.

\textbf{Consistency between the optimal array configurations in paraxial and non-paraxial settings} -- It is instructive to evaluate whether the obtained optimal design conditions in \eqref{eq:4Array_3b} and \eqref{eq:4Array_4b} are consistent with the solution obtained for the paraxial setting in Section \ref{subsec:OrthoParax}. In the paraxial setting, it needs to hold $|\mathbf{c}_o^{i}| \approx |\mathbf{c}_o|$, and hence \eqref{eq:4Array_4b} simplifies as follows:
\begin{equation}
    \frac{M_1^1}{|\mathbf{c}_o|^2} \frac{ 1 - 2|\eta_x^1|}{1 - |\eta_x^1|} \approx - \frac{2M_1^2}{|\mathbf{c}_o|^2} \, .
\end{equation}

By noting that $M_1 = 2M_1^1 + 2M_1^2$, we then obtain $|\eta_x^1| = 1 - M_1^1/M_1$. According to \eqref{eq:4Array_3b}, this provides $|\eta_x^2| = 1/2 - M_1^1/M_1 = M_1^2/M_1$, $\gamma_{11}^{1} = M_1^1/M_1$ and $\gamma_{11}^{2} = M_1^2/M_1$. Inserting these obtained expressions in \eqref{eq:eta_def} and \eqref{eq:gamma_sub}, using again the approximation $|\mathbf{c}_o^{i}| \approx |\mathbf{c}_o|$, and noting that ${\tau}_{11}^{i} \approx 1$ in \eqref{eq:tau_2} for the considered paraxial setting, we obtain the following:
\begin{align}
    &x^{r,1} = \frac{M_1 - M_1^1}{2} \delta_1^{r},  \qquad x^{r,2} =  \frac{M_1^2}{2} \delta_1^{r} \\
    & \delta_1^{r,1} = \delta_1^{r,2} = \delta_1^{r} = \frac{\lambda |\mathbf{c}_o|}{M_1 \delta_1^{t}} \, . \label{eq:orthoCondParax0_FromNonParax}
\end{align}

By direct inspection of \eqref{eq:orthoCondParax0_FromNonParax}, we evince that it coincides with \eqref{eq:orthoCondParax0}, which can be applied in the paraxial setting. This substantiates the consistency of the proposed partioning in sub-arrays in the limiting regime of the paraxial approximation.

\subsubsection{Generic number of sub-arrays}
The approach proposed for sub-arrays with two and four antenna elements can be generalized to the general setting with an arbitrary number (but even for simplicity) of sub-arrays, thus providing a general solution for \eqref{eq:orthoCondULAs3}. 

Specifically, let $N_r$ denote the number of sub-arrays. A general solution for \eqref{eq:orthoCondULAs3} is obtained by first imposing the approximation $f^{i}_-(u-v) \approx - f^{i+1}_+(u-v)$ for $i=1,2,..,{N_r}/{2} - 1$ and $(u - v) = 1,2,...,L_1-1$. This set of equations, results in the following conditions  (for $i=1,2,..,{N_r}/{2} - 1$):
\begin{align}
   & (\gamma_{11}^{i} - |\eta_x^i|) = -(\gamma_{11}^{i+1} + |\eta_x^{i+1}|) \label{eq:NArray_3}\\
  &  \frac{M_1^{i}}{|\mathbf{c}_o^i|^2} \frac{ (\gamma_{11}^{i} - |\eta_x^i|)}{\gamma_{11}^{i}} = - \frac{M_1^{i+1}}{|\mathbf{c}_o^{i+1}|^2} \frac{ (\gamma_{11}^{i+1} + |\eta_x^{i+1}|)}{\gamma_{11}^{i+1}} \label{eq:NArray_4}
\end{align}
provided that $M_1^i / \gamma_{11}^{i} > (L_1 -1)$ for $i=1,2,..,{N_r}/{2}$.

If \eqref{eq:NArray_3} and \eqref{eq:NArray_4} are fulfilled, \eqref{eq:orthoCondULAs3} can then be satisfied by imposing the equality: 
\begin{equation}
\label{eq:orthoCondULAs3_simp}
    f^{1}_+(u-v) + f^{N_r/2}_-(u-v) = 0 \quad \forall u \not= v
\end{equation}
which can in turn be fulfilled by imposing the following conditions:
\begin{align}
    \gamma_{11}^{1} + |\eta_x^1|= 1, \qquad
\gamma_{11}^{N_r/2} - |\eta_x^{N_r/2}| = 0 \, . \label{eq:NArray_2b}
\end{align}

In summary, the proposed design based on the partitioning in sub-arrays is obtained by solving the following system of equations (for $i=1,2,..,{N_r}/{2} - 1$):
\begin{align}
    &\gamma_{11}^{1} + |\eta_x^1|= 1, \quad \gamma_{11}^{N_r/2} - |\eta_x^{N_r/2}| = 0 \\
    & (\gamma_{11}^{i} - |\eta_x^i|) = -(\gamma_{11}^{i+1} + |\eta_x^{i+1}|) \\
    &\frac{M_1^{i}}{|\mathbf{c}_o^i|^2} \frac{ (\gamma_{11}^{i} - |\eta_x^i|)}{\gamma_{11}^{i}} = - \frac{M_1^{i+1}}{|\mathbf{c}_o^{i+1}|^2} \frac{ (\gamma_{11}^{i+1} + |\eta_x^{i+1}|)}{\gamma_{11}^{i+1}}
\end{align}
provided that $M_1^i / \gamma_{11}^{i} > (L_1 -1)$ for $i=1,2,..,{N_r}/{2}$.

In this section, in summary, we have provided three main contributions:
\begin{itemize}
    \item We have introduced an approach for the analysis of LoS MIMO channels in the non-paraxial setting, which is based on a quartic approximation for spherical wavefronts and a sub-array partitioning for large multi-antenna arrays.
    \item Based on the proposed approach, we have introduced an analytical expression for optimizing the positions (in terms of center-points of the sub-arrays and inter-distances between the antenna elements in each sub-array) of the antennas over LoS MIMO channels.
\item We have specialized the proposed design criterion to linear arrays with broadside orientation and have identified explicit analytical expressions for ensuring the orthogonality of the LoS MIMO channel matrix.

\end{itemize}

\section{Numerical Results}\label{sec:NR}
In this section, we present numerical results to validate the proposed analytical framework. The considered setup is the following: $f_c = 28$ GHz ($\lambda=1.07$ cm), $|\mathbf{c}_o| = 256\lambda$, $\alpha = 0$ and $\beta = 0$. As introduced in Section \ref{sec:SM}, the effective rank is utilized as the figure of merit to evaluate the performance of the proposed designs in terms achievable DoF and spatial multiplexing gain.

\begin{figure*}[!t]
     \centering
     \begin{subfigure}[t]{ 0.8\columnwidth}
         \centering
         \includegraphics[width = \columnwidth]{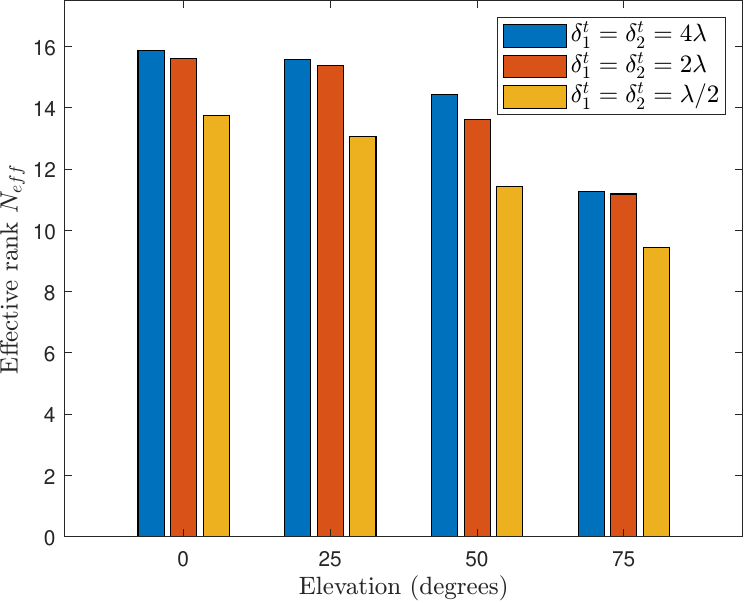}
         \caption{$\delta^r_1$ and $\delta^r_2$ that maximize $N_{\mathrm{eff}}$ using \eqref{eq:channelMatrix}.}
         \label{fig:barplot_exact}
     \end{subfigure}
    \begin{subfigure}[t]{ 0.8\columnwidth}
         \centering
         \includegraphics[width = \columnwidth]{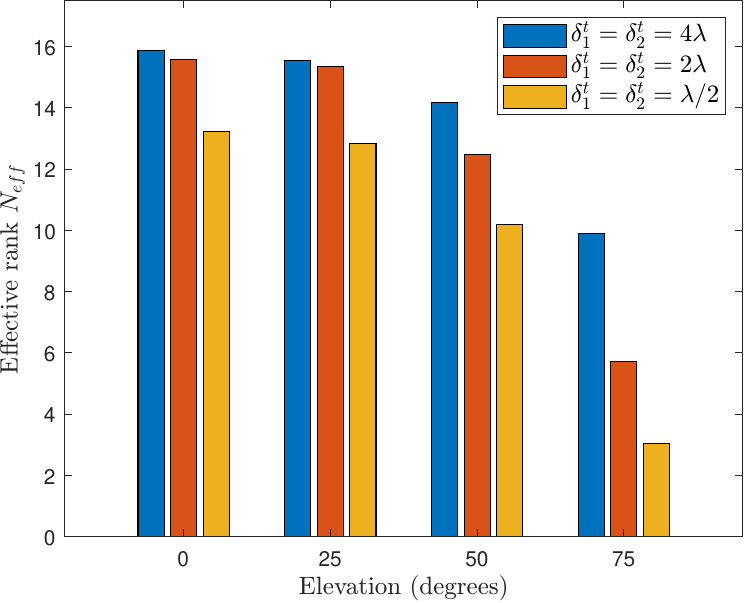}
         \caption{$\delta^r_1$ and $\delta^r_2$ obtained from \eqref{eq:orthoCondParax0}.}
         \label{fig:barplot_num}
    \end{subfigure}
     \caption{Effective rank as a function of the elevation angle (paraxial setting).} 
\end{figure*}
\subsection{Paraxial Setting}
We assume that the multi-antenna transmitter and  receiver are equipped with $4 \times 4$ antenna elements and that they are aligned along the $x$-axis, i.e., $x_o = 0$. The elevation angle of the multi-antenna receiver is $\sin(\theta_o) = z_o/|\mathbf{c}_o|$. The inter-distances $\delta^t_1$ and $\delta^t_2$ at the multi-antenna transmitter are kept fixed, hence the channel orthogonality is obtained by optimizing the inter-distances  $\delta^r_1$ and $\delta^r_2$ at the multi-antenna receiver. 

Figure \ref{fig:barplot_exact} shows the best effective rank that is obtained by optimizing (through an exhaustive grid search) the inter-distances at the receiver based on the exact channel matrix in \eqref{eq:channelMatrix}, and Fig. \ref{fig:barplot_num} shows the effective rank obtained by deploying the antenna elements based on the paraxial design in \eqref{eq:orthoCondParax0}. For low elevation angles, the inter-distances given by the paraxial design provide approximately a full rank channel matrix, i.e., $N_{\mathrm{eff}} \approx 16$. When the elevation angle increases, however, the condition in \eqref{eq:orthoCondParax0} necessitates a larger inter-distance at the multi-antenna receiver, leading to a larger array size. Eventually, the size of the multi-antenna receiver is so large that the paraxial approximation does not hold anymore, resulting in a degradation of the channel orthogonality. Similarly, a shorter inter-distance at the multi-antenna transmitter implies a larger inter-distance at the multi-antenna receiver, resulting in a similar performance degradation. For example, Fig.~\ref{fig:barplot_exact} shows that, in the considered setup, it is not possible to achieve the channel orthogonality for any considered inter-distance at the multi-antenna receiver, when  $\delta^t_1 = \delta^t_2 = \lambda/2$ at the multi-antenna transmitter, which is a typical system design. Thus, the paraxial approximation is not always accurate, and, more importantly, assuming the same inter-distance among all the antenna elements (uniform arrays) is suboptimal even if the effective rank is optimized by using the exact channel in \eqref{eq:channelMatrix}.

\subsection{Non-Paraxial Setting}
In this section, we validate the analytical framework for the non-paraxial deployment. We consider two linear arrays with broadside orientation, i.e., $M_2 = L_2 = 1$, $x_o = 0$ and $z_o = 0$. The transmitter has $L_1 = 16$ antenna elements. To evaluate the accuracy of the analytical framework, we compare four designs to optimize the inter-distances at the multi-antenna receiver:
\begin{itemize}
    \item  \textbf{Design 1}: The multi-antenna receiver is partitioned into four sub-arrays, each having the same number of antenna elements, i.e., $N_r = 4$ and $M_1^i = M_1/4$. The center-points of the sub-arrays are determined from the analytical framework, i.e., by utilizing \eqref{eq:4Array_4d}, \eqref{eq:4Array_3b} and \eqref{eq:eta_def}. The inter-distances in each sub-array are obtained by optimizing (through an exhaustive grid search) the effective rank of the exact channel matrix in \eqref{eq:channelMatrix}.
    \item  \textbf{Design 2}: The same setup as for \textbf{Design 1} is considered, but the channel matrix $\mathbf{H}^{\mathrm{Large}}$ is utilized to maximize the effective rank.
    \item \textbf{Design 3}: The same setup as for \textbf{Design 2} is considered, but the inter-distances at the multi-antenna receiver are obtained from the analytical framework in \eqref{eq:4Array_3b}, \eqref{eq:4Array_4b} and \eqref{eq:gamma_sub}.
    \item \textbf{Design 4}: The multi-antenna receiver is optimized by assuming the paraxial approximation, i.e., the condition in \eqref{eq:orthoCondParax0} is utilized.
\end{itemize}

\begin{figure}[!t]
     \centering
     \includegraphics[width = 0.83\columnwidth]{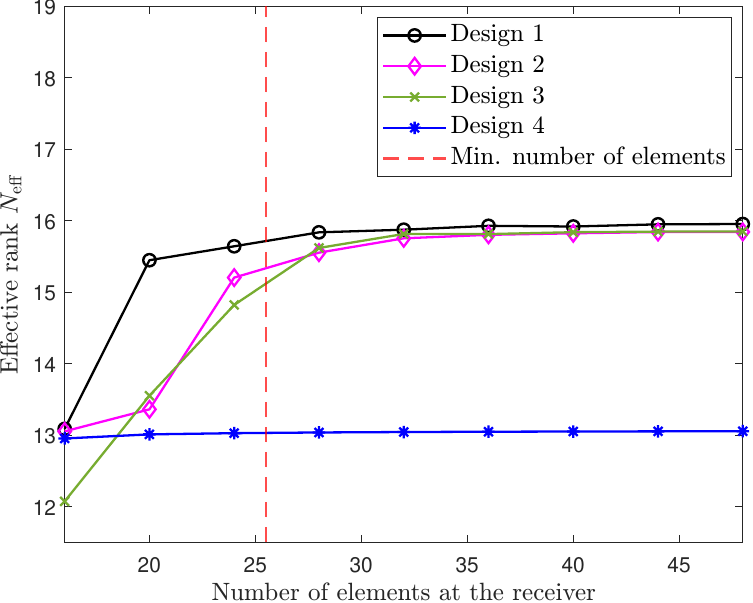}
    \caption{Effective rank versus the number of antennas at the receiver (non-paraxial setting).}
    \label{fig:Eff_R_Nrx} 
\end{figure}

Figure \ref{fig:Eff_R_Nrx} shows the effective rank as a function of the number of antenna elements at the multi-antenna receiver when $\delta_1^t = \lambda/2$. In this setup, the paraxial approximation is not fulfilled, and  Designs 1-3 clearly overcome Design 4 when the configuration for the center-points of the sub-arrays is optimal, i.e., when the number of antenna elements at the receiver is larger than the minimum required (depicted in the figure by a dashed vertical line). According to the example given in Section \ref{sec:BroadsideLinear}, the minimum number of antenna elements at the receiver needs to satisfy the condition $M_1 > 1.7(L_1-1) = 25.5$. It is noteworthy that Design 3, which is based on the proposed analytical framework, results in an effective rank that is similar to that obtained by utilizing numerical grid-based methods, provided that the number of antenna elements at the receiver is sufficiently large, as predicted by the proposed analytical framework.

\begin{figure*}[!t]
     \centering
     \includegraphics[width = \linewidth]{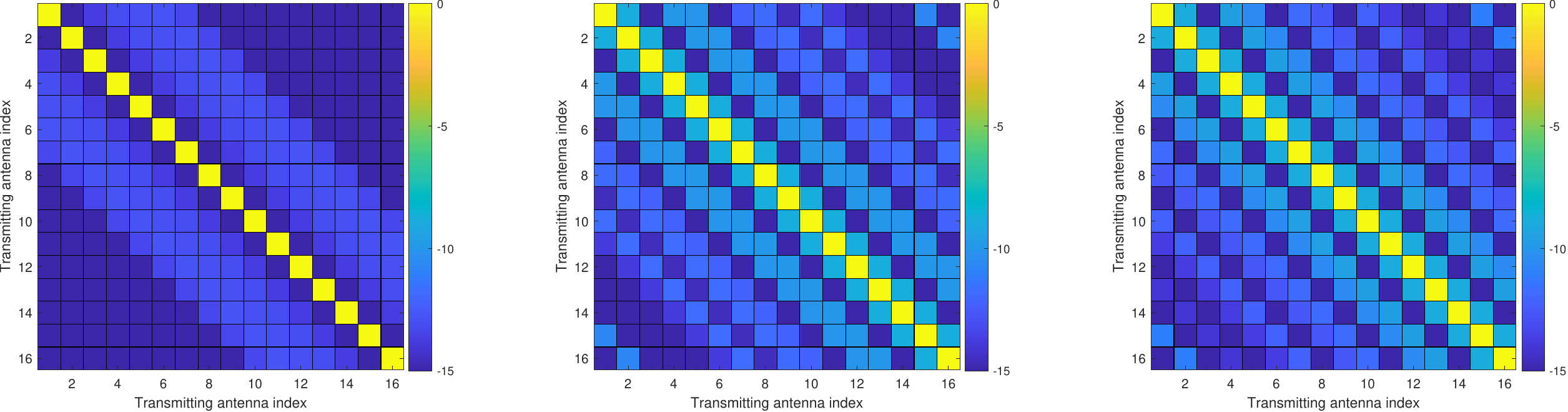}
    \caption{Orthogonality ratio (dB). Design 1 (left), Design 2 (center), Design 3 (right).}
    \label{fig:G_ortho} 
\end{figure*}

Figure \ref{fig:G_ortho} illustrates the matrix $\mathbf{G}$ that is obtained when the multi-antenna receiver is optimized based on Designs 1, 2 and 3 for $M_1 = 48$. We see that the magnitude of the off-diagonal entries of $\mathbf{G}$ is at least 10 dB smaller than the magnitude of the diagonal entries. This confirms the near-orthogonality of the optimized LoS MIMO channel matrix.

\begin{figure}[!t]
     \centering
     \begin{subfigure}[t]{0.8\columnwidth}
         \centering
        \includegraphics[width = \columnwidth]{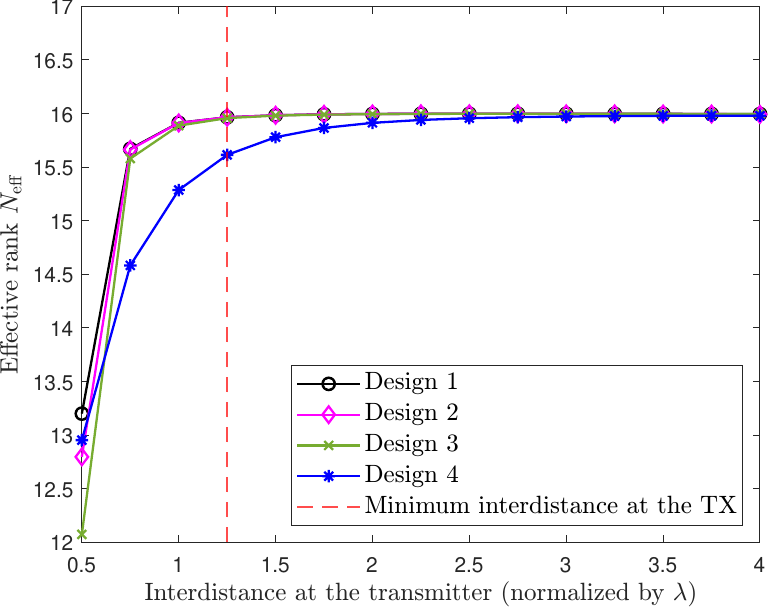}
         \caption{$M_1 = 16$.}
     \end{subfigure}
     \hfill
    \begin{subfigure}[t]{0.8\columnwidth}
         \centering
         \includegraphics[width = \columnwidth]{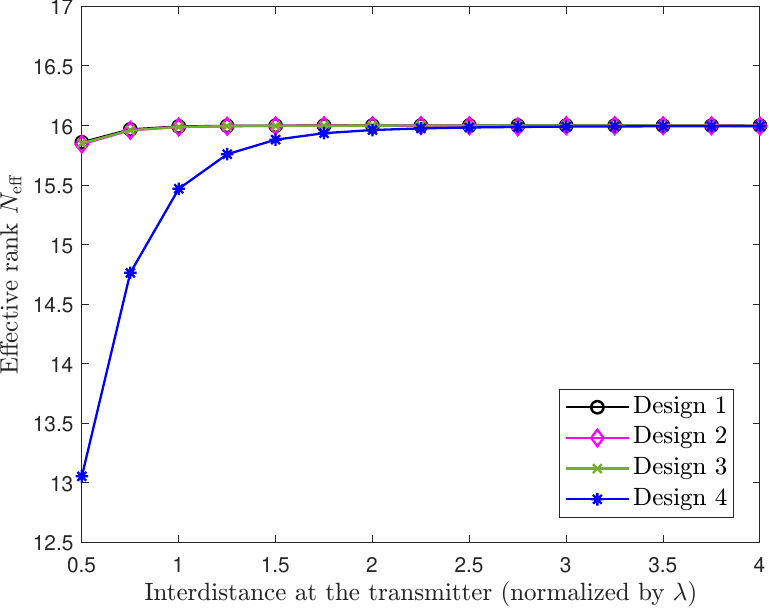}
         \caption{$M_1 = 48$.}
    \end{subfigure}
     \caption{Effective rank as a function of $\delta^t_1$.}
    \label{fig:Neff_deltat}
\end{figure}

Figure \ref{fig:Neff_deltat} shows the effective rank of the LoS MIMO channel as a function of the inter-distance at the multi-antenna transmitter. In this case, the proposed configuration ensures the orthogonality for any inter-distance at the transmitter when $M_1 = 48$ but not when $M_1 = 16$ (the vertical dashed line shows the minimum number of antenna elements based on the proposed framework). Hence, for a given number of antenna elements at the transmitter, there is a minimum required value of the inter-distance at the transmitter for ensuring the orthogonality of the LoS MIMO channel. It needs to be emphasized that the proposed approach offers a sufficient condition to maximize the rank in LoS MIMO channels. Therefore, other designs that ensure that the LoS MIMO channel has a full rank, even when the number of antenna elements at the multi-antenna receiver does not exceed the minimum value estimated in this paper, may exist (as discussed in Section \ref{sec:BroadsideLinear}).

In addition, Fig. \ref{fig:Neff_deltat} shows that, as the inter-distance at the multi-antenna transmitter increases, the inter-distance at the multi-antenna receiver decreases. This is because the paraxial approximation (Design 4) becomes more accurate in this case. Specifically, as shown in Table \ref{tab:interdist}, the inter-distances obtained based on the proposed sub-array partitioning converge towards the inter-distance obtained by considering the paraxial approximation in \eqref{eq:orthoCondParax0}.

\begin{table*}[!t]
\centering
\caption{Inter-distance at the receiver versus the inter-distance at the transmitter ($M_1 = 48$).}
\label{tab:interdist}
\begin{tabular}{|l|l|l|l|l|}
\hline
$\delta^t$ & Design 1 $(\delta^{r,1}, \delta^{r,2})$ & Design 2 $(\delta^{r,1}, \delta^{r,2})$ & Design 3 $(\delta^{r,1}, \delta^{r,2})$ & Design 4 ($\delta^r$) \\ \hline
$0.50\lambda$  & $(63.00\lambda, 20.50\lambda)$     & $(56.00\lambda, 24.50\lambda)$       & $(58.46\lambda, 24.48)$  & $10.67\lambda$           \\ \hline
$\lambda$         & $(6.55\lambda, 5.70\lambda)$       & $(6.30\lambda, 5.85\lambda)$         & $(6.30\lambda, 5.87\lambda)$   & $5.33\lambda$     \\ \hline
$2\lambda$           & $(2.80\lambda, 2.70\lambda)$       & $(2.75\lambda, 2.75\lambda)$         & $(2.77\lambda, 2.72\lambda)$  & $2.67\lambda$      \\ \hline
\end{tabular} 
\end{table*}

\section{Conclusion}\label{sec:Concl}
In this paper, we have introduced a novel framework for optimizing the deployment of the antenna elements in LoS MIMO channels. The proposed approach can be applied to paraxial and non-paraxial settings. In the paraxial setting, we have devised a simple analytical framework that provides explicit expressions for ensuring the orthogonality (i.e., full rank)  of the LoS MIMO channel matrix as a function of key design parameters. In the non-paraxial setting, we have introduced a new analytical framework based on a quartic approximation for spherical wavefronts and the partitioning of large arrays into sub-arrays. The proposed approach provides sufficient conditions for ensuring that the channel matrix is orthogonal, which requires an excess number of antenna elements either at the multi-antenna transmitter or at the multi-antenna receiver. Possible extensions of this paper include the generalization of the proposed methods to deployments with the same number of antenna elements at the transmitter and receiver, the analysis of more complex channel models, e.g., including environmental impairments that affect sub-THz frequencies and multipath interference in urban settings, as well as the impact of possible errors for the optimal positions of the antenna elements at the transmitter and receiver.

\bibliographystyle{IEEEtran}
\bibliography{refs}

\begin{IEEEbiography}[{\includegraphics[width=1in,height=1.25in,clip,keepaspectratio]{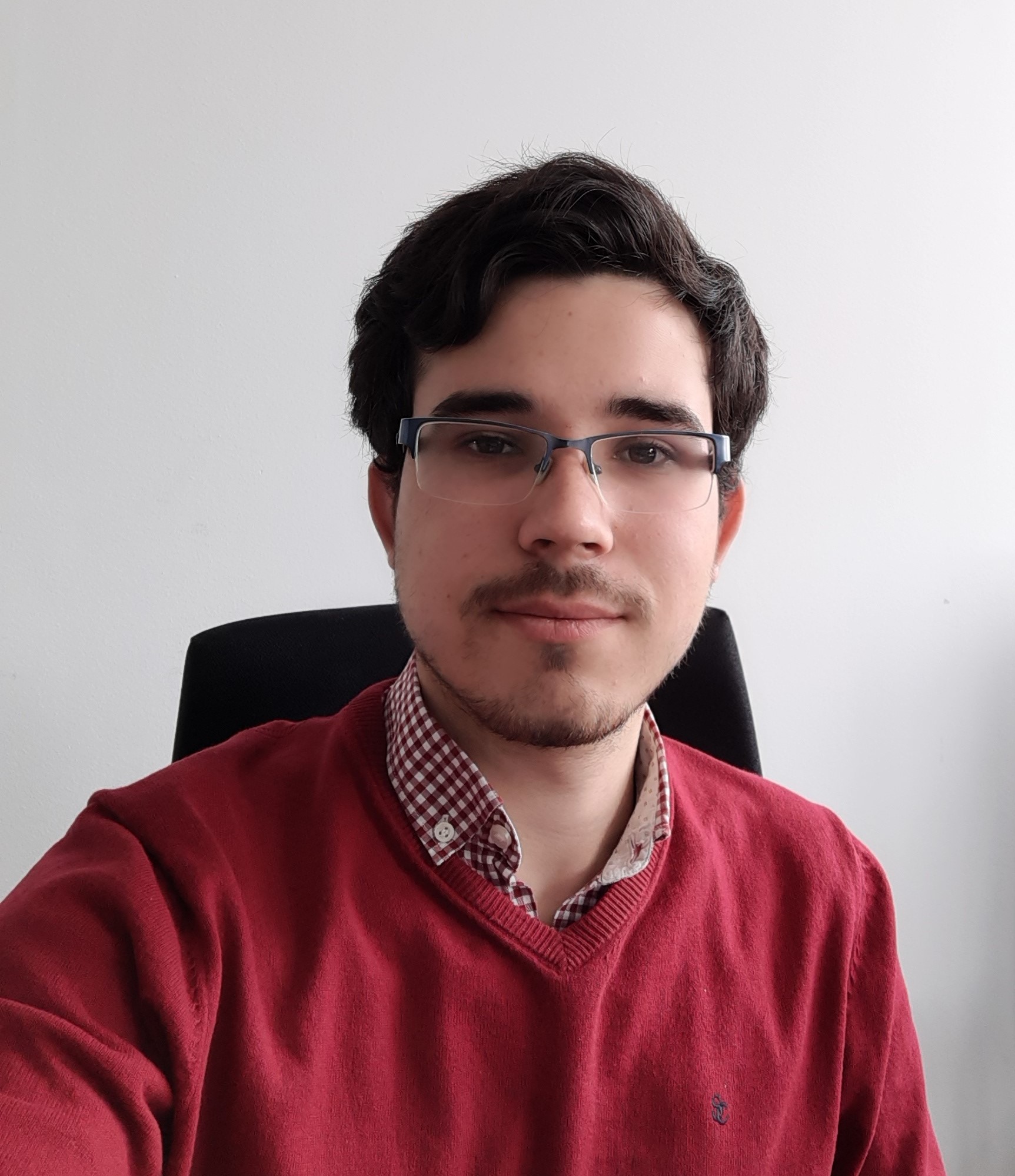}}]{Juan Carlos Ruiz-Sicilia} (Student Member IEEE) received the B.Sc. and M.Sc. degrees in Telecommunication Engineering from the University of Málaga, Spain, in 2019 and 2021, respectively. In 2021, he was part of the German Aerospace Center (DLR), Institute of Communication and Navigation, to carry out his master’s thesis and was the recipient of a national award for his M.Sc. thesis. He is now pursuing a Ph.D. degree in Telecommunications Engineering at Université Paris-Saclay, France. He is employed at CNRS as an Early Stage Researcher in the European project 5GSmartFact H2020 MSCA-ITN.
\end{IEEEbiography}
\begin{IEEEbiography}[{\includegraphics[width=1in,height=1.25in,clip,keepaspectratio]{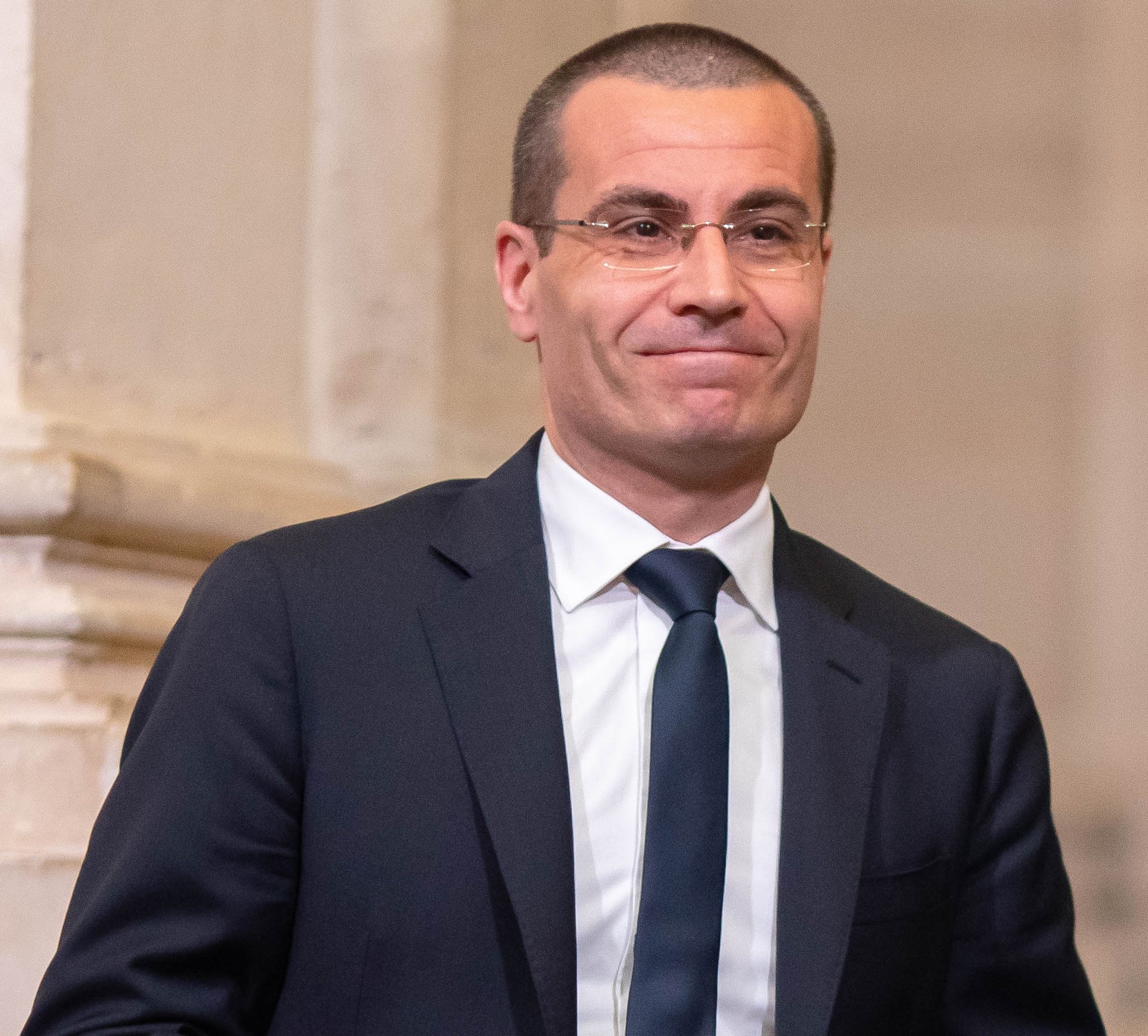}}]{Marco Di Renzo} (Fellow IEEE) received the Laurea (cum laude) and Ph.D. degrees in electrical engineering from the University of L’Aquila, Italy, in 2003 and 2007, respectively, and the Habilitation à Diriger des Recherches (Doctor of Science) degree from University Paris-Sud (currently Paris-Saclay University), France, in 2013. Currently, he is a CNRS Research Director (Professor) and the Head of the Intelligent Physical Communications group in the Laboratory of Signals and Systems (L2S) at Paris-Saclay University -- CNRS and CentraleSupelec, Paris, France. He is a Fellow of the IEEE, IET, EURASIP, and AAIA; an Academician of AIIA; an Ordinary Member of the European Academy of Sciences and Arts, an Ordinary Member of the Academia Europaea; an Ambassador of the European Association on Antennas and Propagation; and a Highly Cited Researcher. Also, he holds the 2023 France-Nokia Chair of Excellence in ICT, he holds the Tan Chin Tuan Exchange Fellowship in Engineering at Nanyang Technological University (Singapore), and he was a Fulbright Fellow at City University of New York (USA), a Nokia Foundation Visiting Professor at Aalto university (Finland), and a Royal Academy of Engineering Distinguished Visiting Fellow at Queen's University Belfast (UK). His recent research awards include the 2022 Michel Monpetit Prize conferred by the French Academy of Sciences, the 2023 IEEE VTS James Evans Avant Garde Award, the 2024 Best Tutorial Paper Award, and the 2024 IEEE COMSOC Marconi Prize Paper Award in Wireless Communications. He served as the Editor-in-Chief of IEEE Communications Letters during the period 2019-2023, and he is currently serving as the Director of Journals of the IEEE Communications Society.
\end{IEEEbiography}
\begin{IEEEbiography}[{\includegraphics[width=1in,height=1.25in,clip,keepaspectratio]{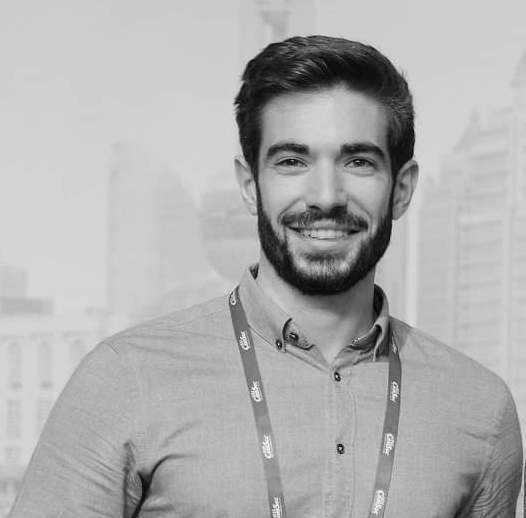}}]{Placido Mursia} (member IEEE) received the B.Sc. and M.Sc. (with honors) degrees in Telecommunication Engineering from Politecnico of Turin in 2015 and 2018, respectively. He obtained his Ph.D. degree from Sorbonne Université of Paris, at the Communication Systems department of EURECOM in 2021. He is currently a Senior Research Scientist with the 6GN group at NEC Laboratories Europe in Heidelberg, Germany. His research interests lie in convex optimization, signal processing, reconfigurable intelligent surfaces, and wireless communications.
\end{IEEEbiography}
\begin{IEEEbiography}[{\includegraphics[width=1in,height=1.25in,clip,keepaspectratio]{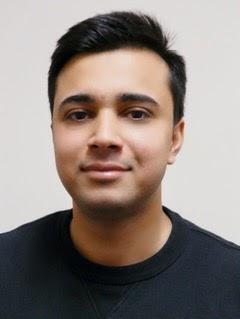}}]{Aryan Kaushik} (Member IEEE) is currently an Assistant Professor at the University of Sussex, UK, since 2021. Prior to that, he has been with University College London, UK (2020-21), University of Edinburgh, UK (2015-19), and Hong Kong University of Science and Technology, Hong Kong (2014-15). He has also held visiting appointments at Imperial College London, UK (2019-20), University of Bologna, Italy (2024), University of Luxembourg, Luxembourg (2018), Athena RC, Greece (2021), and Beihang University, China (2017-19, 2022). He has been External PhD Examiner internationally such as at Universidad Carlos III de Madrid, Spain, in 2023. He has been an Invited Panel Member at the UK EPSRC ICT Prioritisation Panel in 2023 plus Proposal Reviewer for the EPSRC since 2023, and a member of the One6G Association. He is Editor of three upcoming books by Elsevier on Integrated Sensing and Communications, Non-Terrestrial Networks, Electromagnetic Signal and Information Theory, and several journals such as IEEE OJCOMS (Best Editor Award 2023), IEEE Communications Letters (Exemplary Editor 2023), IEEE Internet of Things Magazine (including the AI for IoT miniseries), IEEE Communications Technology News (initiated the IEEE CTN podcast series), and several special issues such as in IEEE Wireless Communications Magazine, IEEE Network Magazine, IEEE Communications Standards Magazine, IEEE Open Journal of the Communications Society, IEEE Internet of Things Magazine.
\end{IEEEbiography}
\begin{IEEEbiography}[{\includegraphics[width=1in,height=1.25in,clip,keepaspectratio]{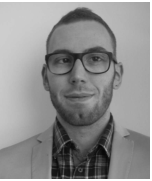}}]{Vincenzo~Sciancalepore} (Senior Member IEEE) received his M.Sc. degree in Telecommunications Engineering and Telematics Engineering in 2011 and 2012, respectively, and a double Ph.D. degree in 2025. He was a recipient of the National Award for the Best Ph.D. Thesis in the area of communication technologies (wireless and networking) issued by GTTI in 2015. Currently, he is a Principal Researcher at NEC Laboratories Europe, focusing his research activity on reconfigurable intelligent surfaces. He is involved in the IEEE Emerging Technologies Committee leading an emerging technology initiative (ETI) on RIS. He is an Editor for IEEE Transactions on Wireless Communications and IEEE Transactions on Communications.
\end{IEEEbiography}

\end{document}